\documentclass[letterpaper]{article} 
\usepackage{aaai2026}  
\usepackage{times}  
\usepackage{helvet}  
\usepackage{courier}  
\usepackage[hyphens]{url}  
\usepackage{graphicx} 
\usepackage{natbib}  
\usepackage{amsmath} 
\usepackage{booktabs}
\usepackage{caption} 
\usepackage{algorithm}
\usepackage{algorithmic}

\usepackage{newfloat}
\usepackage{listings}
\DeclareCaptionStyle{ruled}{labelfont=normalfont,labelsep=colon,strut=off} 
\floatstyle{ruled}
\newfloat{listing}{tb}{lst}{}
\floatname{listing}{Listing}
\title{G-UBS: Towards Robust Understanding of Implicit Feedback \\ via Group-Aware User Behavior Simulation}
\author{Boyu Chen\textsuperscript{\rm1,\rm2,\rm3,\thanks{Equal contribution.}}, Siran Chen\textsuperscript{\rm1,\rm2,\rm3,\footnotemark[1]}, Zhengrong Yue\textsuperscript{\rm6,\footnotemark[1]}, Kainan Yan\textsuperscript{\rm1,\rm2,\footnotemark[1]},
Chenyun Yu\textsuperscript{\rm5, \thanks{Corresponding author.}}, Beibei Kong\textsuperscript{\rm3}, Cheng Lei\textsuperscript{\rm3}, Chengxiang Zhuo\textsuperscript{\rm3}, Zang Li\textsuperscript{\rm3}, Yali Wang\textsuperscript{\rm1,\rm4,\footnotemark[2]}\\
}

\affiliations{

    \rm1 Shenzhen Key Laboratory of Computer Vision and Pattern Recognition, \\ Shenzhen Institutes of Advanced Technology, Chinese Academy of Sciences, Shenzhen, China\\
    \rm2 University of Chinese Academy of Science, Beijing, China\\
    \rm3 Platform and Content Group, Tencent, Shenzhen, China\\
    \rm4 Shanghai Artificial Intelligence Laboratory, Shanghai, China\\
    \rm5 Shenzhen Campus of Sun Yat-sen University, Shenzhen, China\\
    \rm6 Shanghai Jiaotong University, Shanghai, China
    
    {chenboyu18, chensiran17}@mails.ucas.ac.cn, yuchy35@mail.sysu.edu.cn, 
    yl.wang@siat.ac.cn
}

\usepackage{bibentry}
\usepackage{url} 
\usepackage{cite}
\usepackage{amsthm,amsmath,amssymb}
\usepackage{mathrsfs}
\usepackage{array}
\usepackage{tikz}
\usepackage[skins]{tcolorbox} 
\tcbuselibrary{breakable} 
\usepackage{booktabs}  
\usepackage{multirow}
\usepackage{graphicx}
\usepackage{adjustbox}
\usepackage{colortbl}
\usepackage{enumitem} 
\usepackage{bbding} 

\usepackage[capitalize]{cleveref}
\crefname{section}{Sec.}{Secs.}
\Crefname{section}{Section}{Sections}
\Crefname{figure}{Figure}{Figures}
\Crefname{figure}{Fig.}{Figs.}
\Crefname{table}{Table}{Tables}
\crefname{table}{Tab.}{Tabs.}
\crefname{algorithm}{Algorithm}{Algorithms}
\crefname{algorithm}{Alg.}{Algs.}
\crefname{Equation}{Equation}{Equations}
\crefname{Equation}{Eq.}{Eqs.}

\usepackage[caption=false, font=normalsize, labelfont=sf, textfont=sf]{subfig}

\usepackage{etoolbox}
\makeatletter
\patchcmd{\@makecaption}
  {\scshape}
  {}
  {}
  {}
\makeatletter
\patchcmd{\@makecaption}
  {\\}
  {.\ }
  {}
  {}
\makeatother

\usepackage{tabularx} 
\usepackage{ragged2e} 
\newcolumntype{L}{>{\RaggedRight\hangafter=1\hangindent=0em}X}

\usepackage{xspace}
\makeatletter
\DeclareRobustCommand\onedot{\futurelet\@let@token\@onedot}
\def\@onedot{\ifx\@let@token.\else.\null\fi\xspace}

\makeatother


\begin{document}

\maketitle

\begin{abstract}

User feedback is critical for refining recommendation systems, yet explicit feedback (e.g., likes or dislikes) remains scarce in practice. As a more feasible alternative, inferring user preferences from massive implicit feedback has shown great potential (e.g., a user quickly skipping a recommended video usually indicates disinterest). Unfortunately, implicit feedback is often noisy: a user might skip a video due to accidental clicks or other reasons, rather than disliking it. Such noise can easily misjudge user interests, thereby undermining recommendation performance. To address this issue, we propose a novel Group-aware User Behavior Simulation (G-UBS) paradigm, which leverages contextual guidance from relevant user groups, enabling robust and in-depth interpretation of implicit feedback for individual users. Specifically, G-UBS operates via two key agents. First, the User Group Manager (UGM) effectively clusters users to generate group profiles utilizing a ``summarize-cluster-reflect" workflow based on LLMs. Second, the User Feedback Modeler (UFM) employs an innovative group-aware reinforcement learning approach, where each user is guided by the associated group profiles during the reinforcement learning process, allowing UFM to robustly and deeply examine the reasons behind implicit feedback. To assess our G-UBS paradigm, we have constructed a Video Recommendation benchmark with Implicit Feedback (IF-VR). To the best of our knowledge, this is the first multi-modal benchmark for implicit feedback evaluation in video recommendation, encompassing 15k users, 25k videos, and 933k interaction records with implicit feedback. 
Extensive experiments on IF-VR demonstrate that G-UBS significantly outperforms mainstream LLMs and MLLMs, with a 4.0\% higher proportion of videos achieving a play rate \textgreater 30\% and 14.9\% higher reasoning accuracy on IF-VR.

\end{abstract}

\section{Introduction}

Nowadays, multi-modal content platforms like TikTok, Kuaishou, and Tencent Video have become integral to our daily lives. To provide effective and personalized services, these platforms strive to establish accurate recommendation systems by learning user preferences from interaction records, attribute information, and user feedback.
However, when using these platforms, users are rarely active in providing explicit feedback~\cite{xie2021deep} (e.g., ratings, likes, dislikes, and their underlying reasons). Instead, indirect behavioral cues are more observable, such as quick video skips, non-clicks, and low completion rates, which serve as implicit feedback reflecting user discontent.
Consequently, in-depth interpretation of implicit feedback is crucial for boosting recommendation accuracy and personalization.
However, implicit feedback typically contains substantial noise, which can
easily lead to misjudgment of user interests, thereby impairing recommendation performance and ultimately leading to user attrition and platform abandonment~\cite{zhao2023kuaisim}.
For instance, a quick skip may result from accidental operations (e.g., one-handed usage), user habits, or environmental interference, rather than genuine disinterest in the content.
This presents a key challenge: \textit{
how to robustly discern the underlying causes of users’ implicit feedback in the presence of noisy signals?}

Current efforts to mine implicit feedback can be categorized into embedding-based and LLM-based approaches.
Traditional embedding-based approaches \cite{chen2021curriculum, he2016vista, park2022exploiting,guo2017deepfm}
map all implicit feedback into embedding features and feed them into recommendation models under the assumption that richer features will produce better performance. Since these schemes cannot truly capture why users dislike certain content, they tend to result in poor interpretability.
LLM-based approaches leverage large language models (LLMs)~\cite{zhang2025llm,zhang2024usimagent,yang2024oasis} or RL-tuned models~\cite{yang2025reccot,zhao2023kuaisim,zhao2018recommendations} to understand and simulate user behavior (e.g., predicting whether a user will like an item). 
However, existing LLM-based methods focus predominantly on text modality, lacking the ability to jointly perceive information across multiple modalities. In addition, they fail to address the noise in the individual's implicit feedback, further limiting their performance.

\begin{figure}[t]
    \centering
    \includegraphics[width=0.9\linewidth]{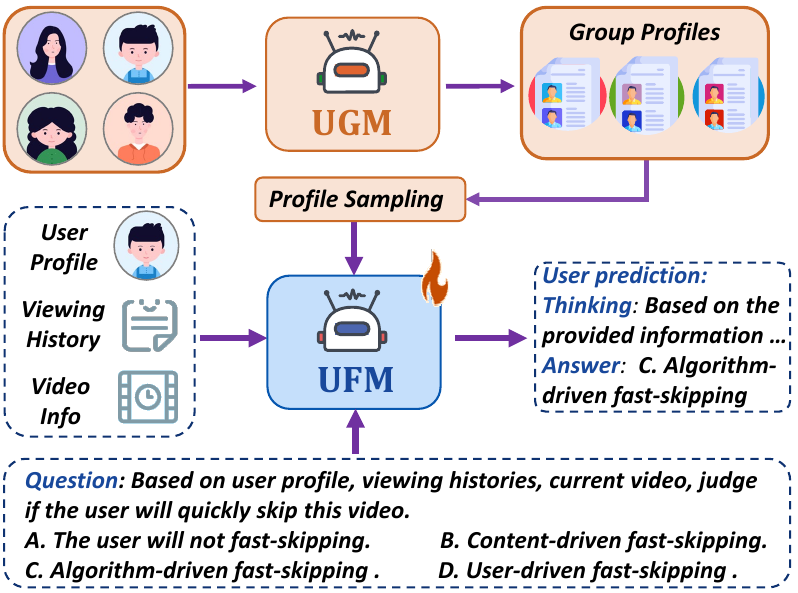}
    \caption{Overview of our G-UBS paradigm: To better visualize implicit feedback, we integrate UGM-generated group profiles into the UFM training process.}   
    \label{fig:intro}
    
\end{figure}

To address these challenges, we propose a novel Group-Aware User Behavior Simulation (\textbf{G-UBS}) paradigm, aiming to robustly and profoundly understand users' implicit feedback under the contextual guidance of relevant user groups as shown in Fig~\ref{fig:intro}.
Specifically,
G-UBS comprises two key collaborative agents, namely User Group Manager (\textbf{UGM}) and User Feedback Modeler (\textbf{UFM}). 
The UGM agent is designed to support 1,000 concurrent users and generate up to 50 distinct group profiles via an LLM-powered ``summarize-cluster-reflect" workflow.
The UFM agent integrates group profiles from UGM and multi-modal information to optimize the training of individual user simulators, effectively filtering noise in implicit negative feedback. 
In summary, G-UBS is a pioneering paradigm that ensures robust scalability while enabling more accurate and reliable simulations to better understand implicit user feedback.

To evaluate our proposed method, we have constructed IF-VR, the first multimodal dataset of user implicit feedback in video recommendation scenario. This dataset covers two mainstream app-based recommendation modes: sequential video recommendation, where users can skip videos they dislike, and click simulation, which predicts click events on pages containing multiple videos and their titles.
Specifically, IF-VR includes data from 15,000 users, covering their demographic profiles and interest tags, along with 933K clicking or watching histories. In addition, it contains 50K dislike feedback entries and 72K annotated types behind users’ fast-swipes or low play rates in viewing histories.
This dataset also includes 25K videos watched by users and their corresponding titles. Sourced from the Tencent Video Mobile app, IF-VR thus stands as a multimodal recommendation dataset that closely aligns with real-world scenarios.
Experimental results demonstrate that our method outperforms other LLMs and MLLMs, excelling in predicting user fast-swipes and non-clicks, with relevant analyses provided. 
Our contributions are summarized below.

\begin{itemize}
\item We propose a novel \textbf{G-UBS} paradigm, which consists of two key agents named \textbf{UGM} and \textbf{UFM}, for integrating group profiles into RL-based user simulation fine-tuning. Our G-UBS eliminates
the noise in the individual implicit feedback with group profile aiding, thereby enhancing the accuracy of user simulation.
    
\item We introduce \textbf{IF-VR}, a large-scale multitask dataset tailored to analyze user implicit feedback in real-world multimodal recommendation scenarios, providing comprehensive evaluation. 
    
\item We have conducted extensive experiments on both real-world business scenarios and open-source datasets. The results demonstrate that our approach outperforms existing LLMs and MLLMs, with a 4.0\% higher proportion of videos achieving a play rate \textgreater 30\% and 14.9\% higher reasoning accuracy.
\end{itemize}

\section{Related Work}

\begin{figure*}[h]
    \centering
    \includegraphics[width=0.9\linewidth]{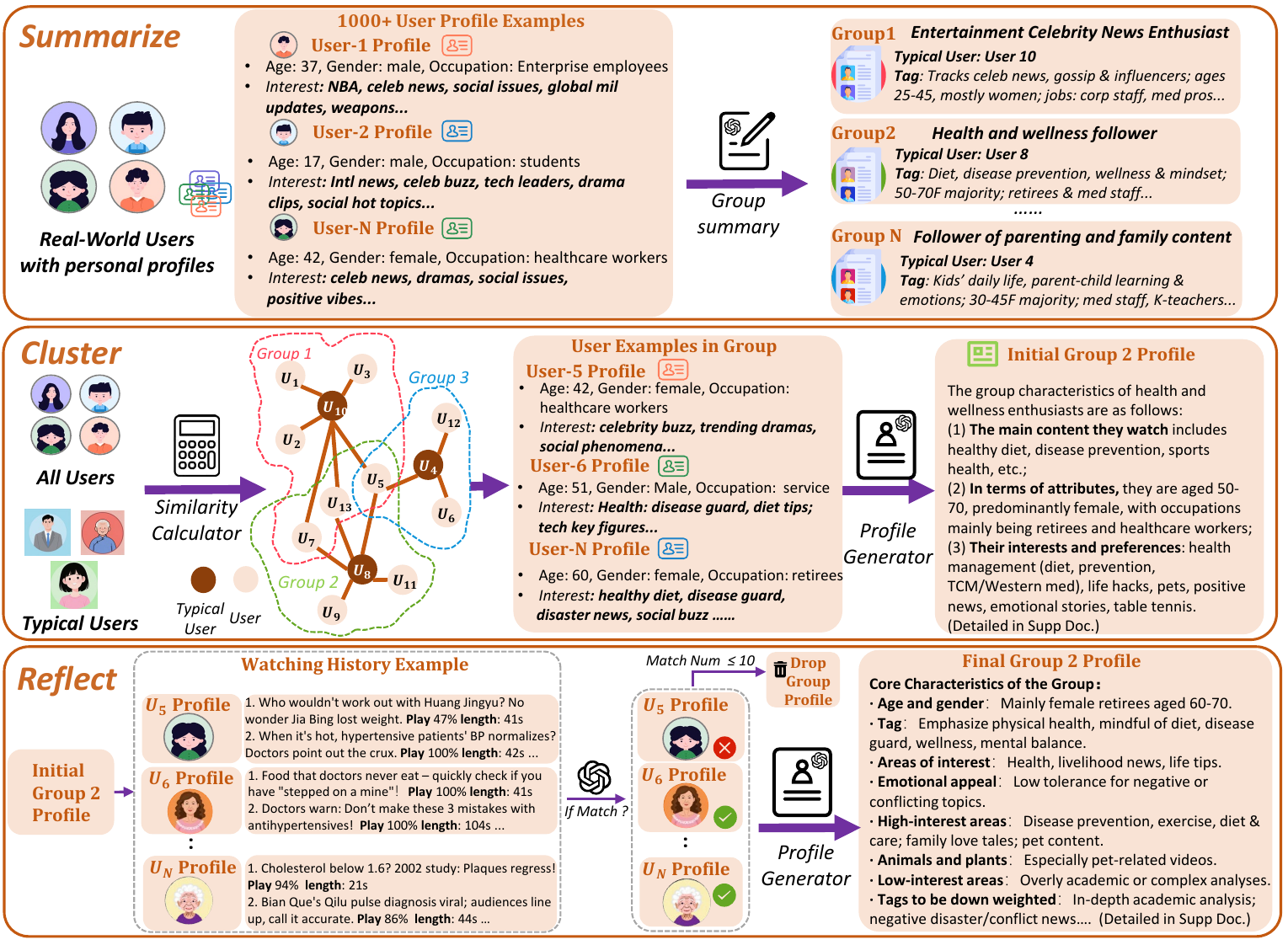}

    \caption{Overview of UGM group profile generation pipeline.}    
    \label{fig:UGM}
    
\end{figure*}

\paragraph{Mining Implicit User Feedback.}

Currently, there are two main paradigms for mining users' implicit feedback: embedding-based and LLM-based methods. Embedding-based methods directly map implicit user feedback to embeddings~\cite{xie2021deep,chen2021curriculum,paudel2018loss,park2022exploiting,lai2024survey}
, which are incorporated into the recommendation pipeline. 
For instance, DFN\cite{xie2021deep} captures unbiased preferences using internal and external feedback, while CDR\cite{chen2021curriculum} uses explicit dislike signals to evaluate behavioral sequences. 
However, such embedding methods suffer from poor interpretability, offering limited insight into the underlying reasons for user dissatisfaction.
LLM-based methods explore the use of LLMs to interpret implicit feedback via CoT reasoning~\cite{yang2025reccot,zhao2025reason,lai2025dar} or the RL training pipeline~\cite{han2025reinforcement,yue2025uniflow,lai2023disentangled}.
However, these single-modal methods do not consider the multimodal feedback noise in individual users.

\paragraph{User Simulator.}
Existing user simulation methods can be divided into two categories. One is system simulation in the context of recommendation systems~\cite{zhang2025llm, zhang2024generative, zhao2023kuaisim, corecco2024suber, zhang2024usimagent, wang2023rethinking,pcaby,videochata1}, which aims to mimic user interactions such as clicks, skips, or feedback. 
However, these single-modal frameworks cannot integrate multimodal signals prevalent in video recommendation scenarios.
The other category is large-scale, LLM-driven multi-agent simulations~\cite{yang2024oasis, piao2025agentsociety, liu2025mosaic,lvagent,sen,vragent,lai2024adaptive}, designed to reproduce emergent behaviors in open-ended social settings and offer insights into collective dynamics. 
However, this large-scale user simulation approach heavily relies on LLMs' capabilities. 
Despite simulating many users, LLMs lack accuracy in understanding users' implicit feedback without fine-tuning.

\begin{figure*}[t]

    \centering

    \includegraphics[width=0.9\linewidth]{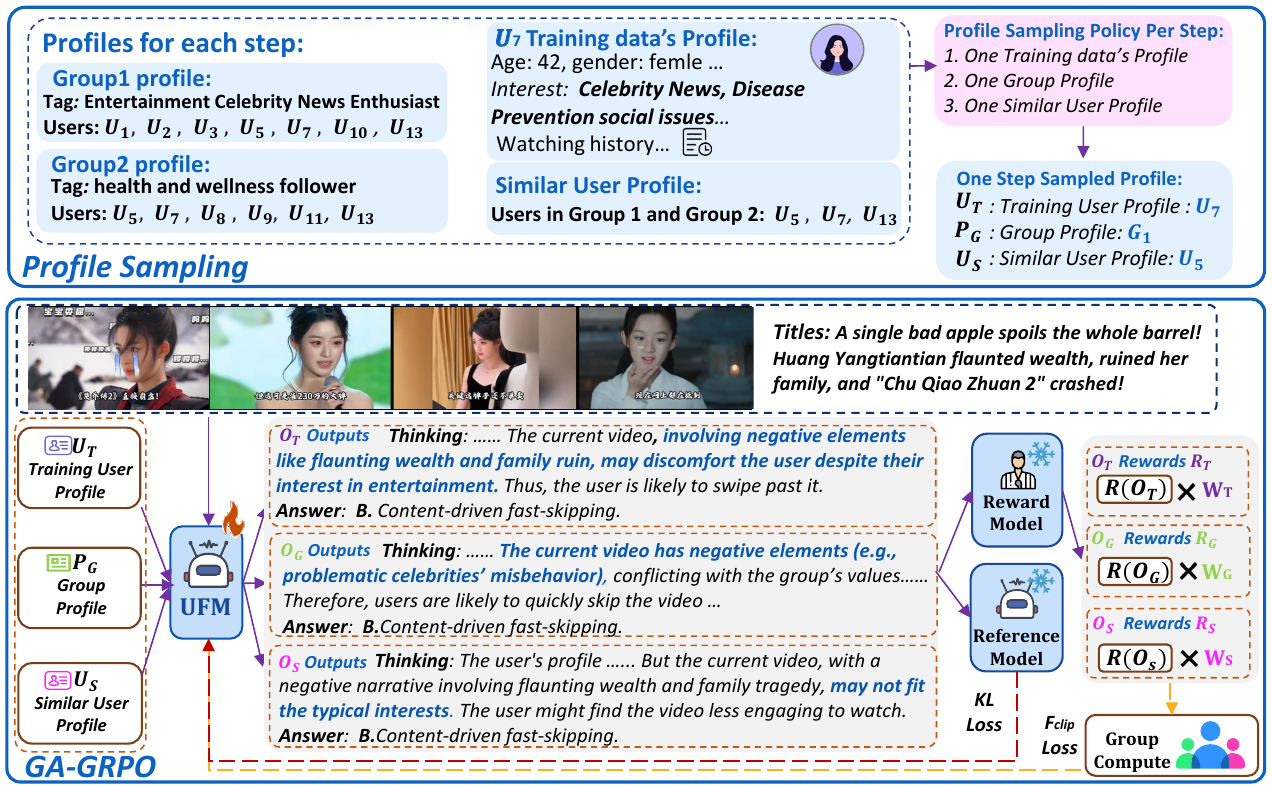}

    \caption{Overview of UFM Reinforcement Learning Paradigm.}    
    \label{fig:UFM}
\end{figure*}

\section{Method}

This section details our Group-Aware User Behavior Simulation (G-UBS) paradigm, where two LLM-based agents (i.e., UGM and UFM) collaborate to integrate group profiles to ensure robust individual user simulations. Then, the construction methodology of the IF-VR dataset is presented.

\subsection{Agent 1: User Group Manager (UGM)}

To obtain robust and representative group profiles while achieving accurate user classification, we propose a summarize-cluster-reflect workflow.
As illustrated in Fig.~\ref{fig:UGM}, UGM first performs initial classification on the user groups to be analyzed and summarizes preliminary group profiles. In the subsequent clustering phase, users are assigned to their respective groups. To mitigate deviations, during the reflection phase, UGM adjusts the prior clusters and outputs refined group profiles. Details are provided as follows. To be specific, the UGM's prompt is in Appendix A.1, detailed initial and final group profiles are in Appendix A.8.

\noindent\textbf{Phase 1: Summarizing Initial Group Profiles.}
To establish an initial classification framework for large-scale user groups to be analyzed (supporting over 1,000 users), we first perform the ``summarize" action.
Given a user profile set $\mathcal{U}$ containing over 1000 users, each user is formatted as $u = [\text{ID}, \text{Occ}, \text{Age}, \text{Gender}, \mathbf{T}],\space u \in \mathcal{U}$.
Here, $\text{Occ}$ denotes the user's occupation and $\mathbf{T}$ represents tags for the user's video preferences as shown in Fig. ~\ref{fig:UGM}. 
The set $\mathcal{U}$, the expected number of categories $k$, and the summary mode $M$ are input into the group summary LLM $\mathcal{S}$ (DeepSeek-R1~\cite{guo2025deepseek}). The model outputs $k$ user groups $\mathcal{G}$ along with their corresponding representative users $U_g$.The mode $M$ specifies the grouping criteria, such as video preferences or demographic attributes (e.g., age, gender, occupation). This process is formulated as:
\begin{equation}
\setlength\abovedisplayskip{3pt}
\setlength\belowdisplayskip{3pt}
U_g, \mathcal{G} =\mathcal{S}(\mathcal{U}, k, M) \quad |\mathcal{G}| = k,  |U_g| = k .
\end{equation}
For each group $g\in\mathcal{G}$, $g =\{T_g, name\}$. $T_g$ is the brief group description tag and $u_g \in U_g$ is the typical user of $g$.

\noindent\textbf{Phase 2: Clustering users to Respective Groups.}
To accurately assign users to their respective groups, in this step, we cluster them by matching each user $u$ to the typical users $u_g \in U_g$ across all categories based on similarity. The top 60 users most similar to each typical user $u_g$ are selected to form the initial user group $C_g$, formulated as:
\begin{equation}
\setlength\abovedisplayskip{3pt}
\setlength\belowdisplayskip{3pt}
\mathcal{C}_g = \left\{ u \in \mathcal{U}, u_g \in U_g \mid \text{Sim}(u, u_g) \geq \tau_g \right\}, g\in\mathcal{G}
\end{equation}
The dynamic threshold $(\tau_j)$ is determined by ranking users in descending order of similarity and selecting the top 60.
These users' profiles are then fed into the profile generator (GPT-4o) to create an initial group profile $\hat{P_g}$, as illustrated in the ``cluster" section of Fig~\ref{fig:UGM}.

\noindent\textbf{Phase 3: Refining Group Profiles via Reflection.}
Considering the context length constraints of LLMs, we did not incorporate user historical viewing records in the previous two working phases, which may lead to misclassifications.
Therefore, in this step, we refine the initial clusters to rectify discrepancies between interest tags and actual viewing behavior.
Taking group $C_g$ as an example: for each user, their profiles $u \in C_g$ and the viewing history $h$ (e.g., play rate, video title, duration, click) are input into the user-matching LLM (GPT-4o), which evaluates whether the user's interest and historical behavior align with $\hat{P_g}$. Users whose preferences match $\hat{P_g}$ are retained in the refined group $ C_g'$.   
\begin{equation}
\setlength\abovedisplayskip{1pt}
\setlength\belowdisplayskip{1pt}
C_g' = \left\{ u \in C_g \mid \text{Match}(u, \hat{P_g}, h) = \text{'Yes'} \right\}
\end{equation}
For each category, if the number of matched users $|C_g'|$ is less than 10, no group profile is generated.
Otherwise, all users' profiles and viewing histories within the group are input into the profile generator (GPT-4o) to summarize key information, including preference characteristics, comprehensive profile details, and recommendation directions, thereby forming the final group profile $P_g$ as illustrated in Fig~\ref{fig:UGM}.

\subsection{Agent 2: User Feedback Modeler (UFM)}

We propose the UFM agent to robustly interpret users' implicit feedback under the guidance of relevant user group profiles, with Reinforcement Learning (RL) employed for model training.
To enable the model to quickly grasp the core logic of the task and enhance the stability of RL, we first perform Supervised Fine-Tuning (SFT) following the approach of DeepSeek-R1~\cite{guo2025deepseek}. Specifically, we leverage 50K collected explicit dislike feedback (e.g., dislike this content, dislike this author) and use GPT-4o~\cite{gpt4o} to generate chain-of-thought annotations for attribution types.
After warming up the model via SFT based on the above explicit dislike signals, we further train UFM following the RL paradigm presented in Fig~\ref{fig:UFM}, which consists of profile sampling and group-aware GRPO (GA-GRPO). Next, we will elaborate on these two components and the reward mechanism.

\begin{table*}[t]
    \centering
    \renewcommand{\arraystretch}{0.7}
    \small
    \setlength{\tabcolsep}{1mm}
    \begin{tabular}{l|ccccccccccc}
    \toprule
    Dataset & Age & Gender & Job &\begin{tabular}[c]{@{}c@{}}Interest\\Tag  \end{tabular} &  \begin{tabular}[c]{@{}c@{}}Video\\Data  \end{tabular}    & \begin{tabular}[c]{@{}c@{}}Explicit\\Feedback  \end{tabular}    &  \begin{tabular}[c]{@{}c@{}}Implicit\\Feedback  \end{tabular}   &  \begin{tabular}[c]{@{}c@{}}Negative\\Comments  \end{tabular}    &  \begin{tabular}[c]{@{}c@{}}Finish\\Rate  \end{tabular}   &  \begin{tabular}[c]{@{}c@{}}Click\\Rate  \end{tabular}    &  \begin{tabular}[c]{@{}c@{}}Play\\Rate  \end{tabular}  \\
    \midrule
    
  Amazon~\cite{hou2024amazon} & \XSolidBrush &\XSolidBrush &\XSolidBrush & \XSolidBrush   &  \XSolidBrush & \Checkmark & \XSolidBrush &  \Checkmark & \XSolidBrush  & \Checkmark   & \XSolidBrush   \\ 
  Netflix~\cite{bennett2007netflix} &\XSolidBrush & \XSolidBrush & \XSolidBrush & \XSolidBrush & \XSolidBrush & \Checkmark & \Checkmark & \XSolidBrush & \Checkmark & \Checkmark   & \XSolidBrush  \\ 
  Yelp~\cite{asghar2016yelp} & \XSolidBrush & \XSolidBrush & \XSolidBrush & \XSolidBrush  &  \XSolidBrush & \Checkmark & \XSolidBrush &  \Checkmark & \XSolidBrush  & \Checkmark   & \XSolidBrush \\ 
  MIND~\cite{wu2020mind} & \XSolidBrush & \XSolidBrush & \XSolidBrush & \Checkmark  &  \XSolidBrush & \XSolidBrush & \Checkmark &  \XSolidBrush & \Checkmark  & \Checkmark   & \XSolidBrush \\ 
  MovieLens~\cite{harper2015movielens} & \Checkmark & \Checkmark & \Checkmark &  \Checkmark     & \XSolidBrush & \Checkmark & \XSolidBrush &  \XSolidBrush & \XSolidBrush & \Checkmark   & \XSolidBrush  \\ 
  MircoLens~\cite{ni2023content} & \XSolidBrush & \XSolidBrush &  \XSolidBrush & \XSolidBrush   &  \Checkmark & \XSolidBrush  & \XSolidBrush & \XSolidBrush & \XSolidBrush & \Checkmark   & \XSolidBrush \\ 
  KuaiRand~\cite{gao2022kuairand} &\XSolidBrush & \XSolidBrush & \XSolidBrush & \XSolidBrush  &  \XSolidBrush & \Checkmark  & \Checkmark & \XSolidBrush & \Checkmark & \Checkmark   & \XSolidBrush \\ \midrule 
  \rowcolor{gray!20}
  IF-VR & \Checkmark & \Checkmark & \Checkmark &  \Checkmark & \Checkmark & \Checkmark & \Checkmark & \Checkmark & \Checkmark  & \Checkmark   & \Checkmark \\\bottomrule

    \end{tabular}
   
    \caption{Comparison with Previous Recommendation Dataset.}
    \label{table:data}
\end{table*}

\noindent\textbf{Profile Sampling.} The UFM RL paradigm includes profile sampling and group-aware GRPO (GA-GRPO) as shown in Fig~\ref{fig:UFM}. GA-GRPO integrates group profiles and similar users into the policy model training, learning from both individual and group-level behaviors to enhance robustness and accuracy. To incorporate group knowledge into individual user simulators, profile sampling is executed at each tuning step, involving three types of profiles: (i) Training User Profile $u_T$: the primary user profile for training.
(ii) Group profile $P_g$: a representative profile of the group to which $u_T$ belongs. If $u_T$ does not belong to any group, $P_g$ is substituted with $u_T$.
(iii) Similar User Profile $u_S$: a profile of another user from the same group(s) as $u_T$. As shown in Fig~\ref{fig:UFM}, if $u_T$ belongs to both Group 1 and Group 2, $u_s$ is sampled from other users in these two groups. If no similar user is available, $u_s$ is replaced with $u_T$. We input the sampled profile to a User Feedback Modeler (UFM) that generates responses $O = \{o_S, o_T, o_G \}$. 
The UFM agent takes the sampled video $V$, its title, a sampled profile $u\in\{u_T, u_S\}$, and user viewing histories $h_T$ and $h_S$ as input, generating outputs $o_T$ and $o_S$ as 
$o = UFM(V, title, u, h)$, where $u\in \{u_T, u_S \}$ and $h\in \{h_T, h_S\}$. For the group profile $P_g$, only the sampled video frames $V$ and title are fed into UFM, yielding $o_G = UFM(V, title, P_g)$.

\noindent\textbf{Reward mechanism.} 
To guide the model in understanding implicit user feedback, we categorize feedback into three types and train UFM using multiple-choice questions. 
To optimiz model performance, we design a reward mechanism that evaluates user behavior predictions based on specific criteria. 
For a given output $o \in O$, three types of rewards are considered:
(i) Format reward: $r_{format}$: to ensure the model generates responses in the desired format (e.g., \texttt{<think>...</think>} for thoughts and \texttt{<answer>...</answer>} for answers), we introduce a format reward $r_{format}$.
(ii) Skip reward: $r_{skip}$: if the model correctly predicts whether the user will fast-forward a video, $r_{skip}$ is assigned.
(iii) Choice reward: $r_{choice}$: if the model accurately chooses the reason options for fast-forwarding, $r_{choice}$ is granted.
The total reward is:
\begin{equation}
\label{eq:out}
\setlength\abovedisplayskip{1pt}
\setlength\belowdisplayskip{1pt}
R(o) = r_{format} + r_{skip} + r_{choice}. 
\end{equation}

\noindent\textbf{Group-Aware GRPO.} To incorporate group-level and similar-user information, we weight the rewards derived from different profiles. Specifically, the rewards $\{R_T, R_S, R_G\}$ are calculated as $R_T = R(o_T) \times W_T$, $R_G = R(o_G) \times W_G$, and $R_S = R(o_S) \times W_S$, respectively. The quality of responses $o \in O$ is evaluated as: 
\begin{equation}
\setlength\abovedisplayskip{3pt}
\setlength\belowdisplayskip{3pt}
\label{eq:ro}
    A_R= 
    \frac{R-\mathrm{mean}(\{R_T, R_S, R_G\})}{\mathrm{std}(\{R_T, R_S, R_G\})}, R\in \{R_T, R_S, R_G\}
\nonumber
\end{equation}
$A_R$ represents the relative quality of $o$. GA-GRPO optimizes for high-scoring responses while incorporating a KL divergence term to constrain the optimized policy $\pi_\theta$ from deviating excessively from the reference policy $\pi_{ref}$, where $\beta$ serves as the regularization coefficient. This process is formulated as:
\begin{equation}
\label{eq:grpo}
\setlength\abovedisplayskip{1pt}
\setlength\belowdisplayskip{1pt}
    \max_{\pi_\theta} \mathbb{E}_{o\sim \pi_{\theta_{\mathrm{old}}}(p)} \Big[
         \sum_{o \in O} \frac{\pi_\theta (o)}{\pi_{\theta_{\mathrm{old}}}(o)} \cdot A_R  - \beta\, \mathrm{D}_\mathrm{KL}\Big(\pi_\theta \,\Vert\, \pi_\mathrm{ref}\Big)
    \Big]
\nonumber
\end{equation}
The Prompt for UFM is in Appendix A.2. Further training details are provided in Appendix A.5.

\subsection{Constructing the IF-VR Dataset}

To address the scarcity of public multimodal datasets specifically designed for attributing users' implicit feedback, we constructed Video Recommendation Dataset with Implicit Feedback (IF-VR) to validate our G-UBS paradigm. IF-VR contains more user information and is closer to real-world recommendation scenarios, as shown in Tab~\ref{table:data}.

\noindent\textbf{Dataset Composition.} As the first multimodal video dataset dedicated to implicit negative feedback verification, IF-VR includes 15K user profiles from the Tencent Video app, with detailed annotations such as gender, age, occupation shown in Tab~\ref{table:data}. The viewing history spans 7 days from May 19 to May 25, 2025.
This dataset covers two mainstream app-based recommendation modes:
(i) Sequential video recommendation, allowing to skip disliked videos, with 8,000 users, 320K viewing histories, 50K explicit "dislike" feedback, and 72K implicit feedback annotations generated by GPT-4o and checked by humans.
(ii) Click simulation, predicting user clicks on pages with multiple videos and titles, including 7,000 users, 613K exposure/click histories. In summary, IF-VR encompassed 15k users, 25K videos, and 993K interaction records. Appendix A.4 shows more details.

\begin{table*}[t]
    \centering
    \small
    \renewcommand{\arraystretch}{0.8}
    \begin{tabular}{l|cccc|c|cccc}
    \toprule
    Model & \begin{tabular}[c]{@{}c@{}}Person\\Play Rate  \end{tabular}   & \begin{tabular}[c]{@{}c@{}}Total\\Play Rate  \end{tabular}   & \begin{tabular}[c]{@{}c@{}}Finish\\Rate  \end{tabular}   & \begin{tabular}[c]{@{}c@{}}Play Rate\\ \textgreater 30\%  \end{tabular} & \begin{tabular}[c]{@{}c@{}}Click\\Rate \end{tabular} & \begin{tabular}[c]{@{}c@{}}Judge\\F1  \end{tabular} & \begin{tabular}[c]{@{}c@{}}Judge\\Acc  \end{tabular} & \begin{tabular}[c]{@{}c@{}}Reason\\F1  \end{tabular} & \begin{tabular}[c]{@{}c@{}}Reason\\Acc  \end{tabular}  \\ 
    \midrule
    Original Recommendation & 46.5\% & 48.3\% & 17.1\% & 76.3\% & 21.4\% & - & - & - & - \\ 
    SASRec~\cite{kang2018self} & 45.8\% & 48.0\% & 16.9\% & 75.8\% & 20.9\%& - & - & - & - \\ 
    
    \midrule

    Llama3.3-70b~\cite{grattafiori2024llama} & 46.5\% & 48.4\% & 17.3\% & 77.0\% & 21.6\% & 18.1\% & 71.9\% & 10.3\% & 8.6\% \\ 
    Qwen3-32b~\cite{yang2025qwen3} & 47.8\% & 49.6\% & 17.6\% & 80.0\% & 21.8\% & 36.0\% & 64.1\% & 36.5\% & 35.1\% \\ 
    Qwen3-235b~\cite{yang2025qwen3} & 48.3\% & 51.6\% & 18.3\% & 83.8\% & 21.9\% & 44.3\% & 65.9\% & 38.6\% & 42.3\% \\ 
    Deepseek-r1-0528~\cite{guo2025deepseek} & 49.6\% & 53.0\% & 20.9\% & 83.8\% & 22.7\% & 43.0\% & 61.0\% & 41.3\% & 48.0\%  \\ 
    \midrule
    Qwen-2.5VL-7B~\cite{yang2024qwen2} & 47.2\% & 49.0\% & 17.1\% & 79.3\% & 22.0\% & 35.8\% & 55.6\% & 30.6\% & 36.4\% \\ 
    Video-R1~\cite{video-r1} & 48.3\% & 49.1\% & 17.6\% & 80.4\% & 22.2\% & 36.7\% & 57.9\% & 31.7\% & 36.9\% \\ 
    Videochat-R1~\cite{li2025videochat-r1} & 48.4\% & 52.5\% & 19.7\% & 84.5\% & 22.1\% & 42.4\% & 42.9\% & 42.0\% & 41.6\% \\ 
    Doubao-1.5-pro~\cite{doubao} & 48.8\% & 50.0\% & 18.4\% & 81.0\% & 22.8\% & 38.6\% & 70.5\% & 16.5\% & 22.5\% \\ 
    GPT-4o~\cite{gpt4o} & 51.3\% & 52.8\% & 19.9\% & 84.7\% & 23.0\% & 45.4\% & 65.1\% & 37.4\% & 40.5\% \\ 
    \midrule
    \rowcolor{gray!20}
    \textbf{G-UBS} & \textbf{52.3\%} & \textbf{55.3\%} & \textbf{22.1\%} & \textbf{88.7\%} & \textbf{25.7\%} & \textbf{54.9\%} & \textbf{72.9\%} & \textbf{55.6\%} & \textbf{62.9\%} \\ \bottomrule
    \end{tabular}
    \caption{SOTA Comparison on Different LLM and MLLM in Understanding the Implicit Feedback of Users}
    \label{sota_table}
\end{table*}

\begin{table}[!]
\centering
\small
\setlength{\tabcolsep}{1.2mm}
\begin{tabular}{lccccccc}
\toprule
 & \multicolumn{3}{c}{\textbf{MovieLens}} & \multicolumn{3}{c}{\textbf{Amazon Books}} \\ 
\cmidrule(lr){2-4} \cmidrule(lr){5-7}
\textbf{Method} & \textbf{Acc} & \textbf{Recall} & \textbf{F1} & \textbf{Acc}  & \textbf{Recall} & \textbf{F1} \\ \midrule

RecAgent\ & 58.1\%  & 60.4\% & 62.1\% & 62.7\%  & 64.9\% & 65.0\% \\ 

Agent4Rec  & 69.1\%  & 69.1\% & 69.8\% & 68.9\%  & 70.3\% & 67.9\% \\

GPT-4o & 72.2\% & 71.8\% &  73.6\%  &  73.4\% &  72.8\%  & 73.6\%  \\ 

SimUser  & 79.1\% & 75.8\% & 77.7\% & 79.1\%  & 78.5\% & 79.4\% \\
 \rowcolor{gray!20}
\textbf{G-UBS} &  \textbf{79.9\%} & \textbf{76.2\%} & \textbf{78.2\%} & \textbf{80.1\%} &\textbf{ 78.9\%} & \textbf{80.2\% }  \\

\bottomrule
\end{tabular}
\caption{User Simulation Experiment on Public Datasets.}
\label{table:taste_alignment}
\end{table}

\begin{table}[!]
    \centering
    \small
    \begin{tabular}{c|ccccc}
    \toprule
    \begin{tabular}[c]{@{}c@{}}Group\\Num  \end{tabular} & \begin{tabular}[c]{@{}c@{}}Person\\Play Rate  \end{tabular}    & \begin{tabular}[c]{@{}c@{}}Play Rate\\ \textgreater 30\% \end{tabular} & \begin{tabular}[c]{@{}c@{}}Click\\Rate \end{tabular}  & \begin{tabular}[c]{@{}c@{}}Judge\\ F1  \end{tabular}  & \begin{tabular}[c]{@{}c@{}}Reason\\ F1  \end{tabular} \\ 
    \midrule
    10  &  52.1\%   &   88.5\%   &  25.5\%   & 54.5\% & 55.1\% \\ 
    \rowcolor{gray!20}
    \textbf{20} &  \textbf{52.3\%}  &  \textbf{88.7\%} &  \textbf{25.7\%}    &  \textbf{54.9\% }  &  \textbf{55.6\%} \\ 
    30 &  52.3\%  &  88.6\%    &  25.4\%  & 54.6\% & 55.3\%  \\ 
    40 &  52.3\%   &  88.7\%   &  25.2\%   &  54.3\% & 55.4\% \\ \bottomrule
    \end{tabular}
    \caption{Different Grouping Numbers on UGM}
    \label{class_num}
    
\end{table}

\begin{table}[t]
\centering
    \small
    \setlength{\tabcolsep}{0.5mm}
\renewcommand{\arraystretch}{0.75}
    
    \begin{tabular}{cc|ccccc}
    \toprule
    Interest & Demographics & \begin{tabular}[c]{@{}c@{}}Person\\Play Rate  \end{tabular}    & \begin{tabular}[c]{@{}c@{}}Play Rate\\ \textgreater 30\% \end{tabular} & \begin{tabular}[c]{@{}c@{}}Click\\Rate \end{tabular}  & \begin{tabular}[c]{@{}c@{}}Judge\\ F1  \end{tabular}  & \begin{tabular}[c]{@{}c@{}}Reason\\ F1  \end{tabular}\\ 
    \midrule

    \rowcolor{gray!20} 
    \Checkmark &  \XSolidBrush & \textbf{52.3\%}  &  \textbf{88.7\%} &     \textbf{25.7\%} &  \textbf{54.9\%}   &  \textbf{55.6\%}\\  
    \XSolidBrush &  \Checkmark  &   52.0\%  &  88.3\%   &  25.6\%    &  54.6\% &  55.1\%  \\ 
    \Checkmark &  \Checkmark     & 52.2\%   & 88.7\%    &  25.4\%   &  54.8\% & 55.3\% \\ 
    
    \bottomrule
    \end{tabular}
    \caption{Different Grouping Methods on UGM}
    \label{catigration_method}
\end{table}

\begin{table}[!]
    \small
    \renewcommand{\arraystretch}{0.9}
    \setlength{\tabcolsep}{1.8mm}
    \centering
    \begin{tabular}{c|ccccc}
    \toprule
    Model & \begin{tabular}[c]{@{}c@{}}Person\\Play Rate  \end{tabular}    & \begin{tabular}[c]{@{}c@{}}Play Rate\\ \textgreater 30\% \end{tabular} & \begin{tabular}[c]{@{}c@{}}Click\\Rate \end{tabular}  & \begin{tabular}[c]{@{}c@{}}Judge\\ F1  \end{tabular}  & \begin{tabular}[c]{@{}c@{}}Reason\\ F1  \end{tabular} \\ 
    \midrule
    BERT-Sim &  51.8\% &  88.0\%   & 24.9\%    & 54.3\%    &  54.7\% \\ 
    K-Means & 51.6\%  & 87.9\%    &   25.0\%  &  54.6\%   & 55.0\% \\ 
    \rowcolor{gray!20}
    TFIDF & \textbf{52.3\%}  &  \textbf{88.7\%} &  \textbf{25.7\%}    &  \textbf{54.9\%}   &  \textbf{55.6\%} \\ 
     \bottomrule
    \end{tabular}
    \caption{Different User Clustering Methods on UGM}
    \label{clustering_user}
\end{table}

\begin{table}[!]
    \small
    \setlength{\tabcolsep}{1mm}
\renewcommand{\arraystretch}{0.6}
    
    \begin{tabular}{ccc|ccccc}
    \toprule
    SFT & RL & \begin{tabular}[c]{@{}c@{}}Group\\Profile  \end{tabular}  & \begin{tabular}[c]{@{}c@{}}Person\\Play Rate  \end{tabular}    & \begin{tabular}[c]{@{}c@{}}Play Rate\\ \textgreater 30\% \end{tabular} & \begin{tabular}[c]{@{}c@{}}Click\\Rate \end{tabular}  & \begin{tabular}[c]{@{}c@{}}Judge\\ F1  \end{tabular}  & \begin{tabular}[c]{@{}c@{}}Reason\\ F1  \end{tabular}\\ 
    \midrule
     \XSolidBrush & \XSolidBrush & \XSolidBrush & 47.2\%   &   79.3\%  &   22.8\%  &  35.8\%   & 30.6\%\\ 
     \Checkmark & \XSolidBrush & \XSolidBrush & 48.6\%   &   80.8\%  &   23.6\%  &   38.0\%  & 36.4\%\\ 
     \XSolidBrush & \Checkmark & \XSolidBrush &  49.6\%  &  84.7\%   &  23.2\% & 45.3\%  &  46.0\%   \\ 
     \Checkmark & \Checkmark & \XSolidBrush &  51.2\%  &  87.4\%   &   23.8\%  &  51.4\%  & 46.5\%\\ 
     \XSolidBrush & \Checkmark & \Checkmark &  50.8\%  & 87.9\%    &  24.9\%   & 52.6\%     & 50.0\%\\ 
     \rowcolor{gray!20}
    \Checkmark & \Checkmark & \Checkmark &  \textbf{52.3\%}  &  \textbf{88.7\%}   &  \textbf{25.7\%}   & \textbf{54.9\%}    & \textbf{55.6\%} \\ \bottomrule
    \end{tabular}

    \caption{Training Processes on UFM}
    \label{whether_train}
    
\end{table}

\noindent\textbf{Labeling for IF-VR.} To attribute these users' implicit feedback, we categorized the underlying causes into three types:
\textbf{(i) Content-driven fast-skipping} (due to objective content flaws): e.g., vulgar content, clickbait titles, physiologically discomforting visuals (e.g., bloody scenes, unpleasant creatures like snakes or centipedes).
\textbf{(ii) Algorithm-driven fast-skipping} (due to recommendation ineffectiveness): e.g., inaccurate user profiling, repetitive recommendations, insufficient diversity.
\textbf{(iii) User-driven fast-skipping} (arising from individual user actions): such as operational errors and lack of viewing intent at the current time.
To perform this categorization, we instructed GPT-4o to label the 72K histories with a viewing rate below 0.3 (from the 320K viewing entries) using the three types mentioned above, and double-checked by humans. More details are in Appendix A.3.

\section{Experiments}

\paragraph{Implementation Details and Metrics.} 
In our experiments, we utilize Tencent Video APP's native recommendation system to generate a candidate video set, then apply our proposed G-UBS paradigm to filter out items mismatched with user preferences, yielding more precise outcomes. The LLM employed throughout the G-UBS workflow is Qwen2.5-VL-7B~\cite{bai2025qwen2}. We perform full-parameter fine-tuning on the UFM agent using 4 A100 80G GPUs.
For both SFT and RFT, the learning rate is set to 1e-5. We conducted SFT for 1 epoch and RFT for 200 steps to achieve optimal results on IF-VR. Additional training details and experimental settings for IF-VR, MovieLens, and Amazon Books are provided in Appendix A.6 and A.7.
The evaluation metrics in Tab~\ref{sota_table} are defined as follows: Play Rate is the ratio of watch time to video duration, with repeated views counted as 100\%. Person Play Rate and Total Play Rate refer to the average Play Rate per user and across all videos, respectively. Finish Rate is the proportion of videos with Play Rate\textgreater90\%, and Play Rate\textgreater30\% is the percentage of videos with a Play Rate\textgreater30\%. Click Rate is the ratio of clicks to recommended videos. Judge F1/Acc evaluate the model's ability to predict skips, while Reason F1/Acc assess its accuracy in identifying why users skip a video.
Following SimUser~\cite{xiang2024simuser}, the metrics (ACC, Recall, F1) in Tab~\ref{table:taste_alignment} measure the model's ability to predict whether users will like a video.

\begin{figure*}[t]
    \centering
    \includegraphics[width=1\linewidth]{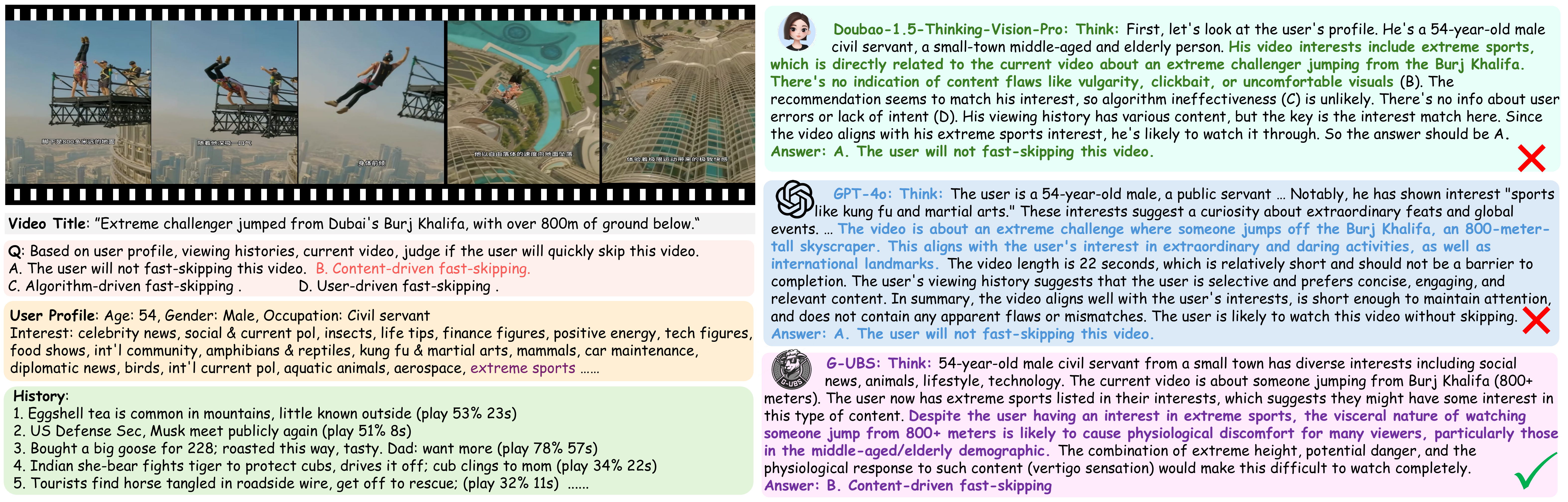}
    \caption{Case Study of G-UBS.}    
    \label{fig:visual}
    
\end{figure*}

\paragraph{Comparison with SOTAs.}
As shown in Tab~\ref{sota_table}, we compare our method with Tencent Video APP's original recommendation strategy,
methods based on LLMs~\cite{grattafiori2024llama,yang2025qwen3,guo2025deepseek} and MLLMs~\cite{yang2024qwen2,video-r1,doubao,gpt4o,li2025videochat-r1}, and the traditional item-based SASRec~\cite{kang2018self}. 
Experimental results show that our post-processed recommendation strategy achieves the best overall performance.
Traditional methods, such as SASRec, rely solely on item embeddings and fail to grasp concrete item semantics. Meanwhile, LLMs and MLLMs poorly align with user preferences due to inherent noise in user feedback.
Specifically, our method significantly boosts user engagement: Person Play Rate increases from 46.5\%  to 52.3\%, and Total Play Rate rises from 48.3\% to 55.3\%, outperforming top LLM (Deepseek-R1) and MLLM (GPT-4o) baselines, attributing to learning from different users within groups.
Moreover, our model achieves the highest Reasoning Accuracy and F1 scores, indicating a strengthened capability to interpret users' implicit feedback. These results confirm the G-UBS paradigm's effectiveness.

\paragraph{User Simulation Experiments on Public Datasets.}

To verify G-UBS's applicability, we compared it with RecAgent~\cite{wang2025user}, Agent4Rec~\cite{zhang2024agent4rec}, GPT-4o~\cite{gpt4o}, and SimUser~\cite{xiang2024simuser} using MovieLens and Amazon Books dataset. Following SimUser~\cite{xiang2024simuser}'s approach, we selected 1000 user profiles from public datasets. For each user, 20 items were randomly chosen (10 clicked and 10 unclicked by the user). The user simulator agent must clearly understand the user's preferences. The G-UBS and other LLM/MLLM models are required to classify these items as liked or disliked. Tab~\ref{table:taste_alignment} reports the binary classification accuracy of this experiment, which measures the model's ability to simulate user click responses. Our G-UBS achieves the highest simulation accuracy on both public datasets.



\subsection{Ablation Study}

\noindent\textbf{Different Grouping Methods on UGM.}
We ablate the optimal grouping strategies for the `summarize' process as shown in Tab~\ref{catigration_method}. Interest-based grouping outperforms both demographic-only and hybrid (interest+demographic) approaches. This is because demographic grouping (by age, gender, or occupation) suffers from large intra-group interest divergence. Users sharing the same attributes often have different video preferences and opinions, reducing the consistency of group profiles.

\noindent\textbf{Different User Clustering Methods on UGM.}
We conduct various clustering methods to determine the optimal choice in the `cluster' process of UGM. In Tab~\ref{clustering_user}, TF-IDF outperforms BERT and traditional ML methods. Since user profiles are structured as concatenated word sequences rather than natural language texts, TF-IDF outperforms BERT (which excels at sentence understanding) and K-Means (which lacks semantic understanding) in word matching.

\noindent\textbf{Training Processes on UFM.}
We conduct ablation studies on the UFM training process(with Qwen2.5-VL as the baseline) to analyze the impacts of SFT, RL, and group profile. As shown in Tab~\ref{whether_train}, SFT and RL boost performance incrementally. But adding group profiles, especially SFT+RL+Group, drives the best results. This result validates the efficacy of our UFM agent tuning pipeline.

\subsection{Hyperparameter Analysis}

\noindent\textbf{Different Grouping Numbers on UGM.}
We vary the number of groups in the `summarize' process of UGM to determine the optimal granularity. The results from Table~\ref{class_num} shows that 20 groups yield the best performance (e.g., 52.3\% Person Play Rate). Fewer groups lead to significant intra-group interest divergence, which undermines the consistency of group profiles. Conversely, an excessive number of groups reduces the number of users per group, resulting in group profiles that only serve a small minority and lack representativeness, thereby leading to suboptimal results.

\noindent\textbf{Visualization of G-UBS.} 
Regarding the video of jumping off the Burj Khalifa in Fig.\ref{fig:visual}, both Doubao and GPT-4o concluded that the user would not fast-skip it based on the user's profile. However, such high-altitude extreme sports videos can cause dizziness and physical discomfort, especially for middle-aged and elderly people. Our G-UBS can accurately determine why the user would fast-skip it based on the group characteristics of middle-aged and elderly people.

\section{Conclusion}
In this paper, we propose a novel Group-aware User Behavior Simulation (G-UBS) paradigm that captures and leverages group profiles to achieve a more robust and profound understanding of users’ implicit feedback.
We have also constructed a large-scale IF-VR dataset to support experimental evaluations. 
Compared with mainstream LLMs and MLLMs, G-UBS achieves a 4.0\% higher proportion of videos with a play rate exceeding 30\% and a 14.9\% improvement in reasoning accuracy on the IF-VR dataset. Extensive experiments validate the efficacy of our proposed method.

\section{Acknowledgement}
This work was supported by the National Key R\&D Program of China (NO.2022ZD0160505).

\bibliography{11_reference}

@inproceedings{zhang2025llm,
  title={Llm-powered user simulator for recommender system},
  author={Zhang, Zijian and Liu, Shuchang and Liu, Ziru and Zhong, Rui and Cai, Qingpeng and Zhao, Xiangyu and Zhang, Chunxu and Liu, Qidong and Jiang, Peng},
  booktitle={Proceedings of the AAAI Conference on Artificial Intelligence},
  pages={13339--13347},
  year={2025}
}

@article{grattafiori2024llama,
  title={The llama 3 herd of models},
  author={Grattafiori, Aaron and Dubey, Abhimanyu and Jauhri, Abhinav and Pandey, Abhinav and Kadian, Abhishek and Al-Dahle, Ahmad and Letman, Aiesha and Mathur, Akhil and Schelten, Alan and Vaughan, Alex and others},
  journal={arXiv preprint arXiv:2407.21783},
  year={2024}
}

@article{bai2025qwen2,
  title={Qwen2. 5-vl technical report},
  author={Bai, Shuai and Chen, Keqin and Liu, Xuejing and Wang, Jialin and Ge, Wenbin and Song, Sibo and Dang, Kai and Wang, Peng and Wang, Shijie and Tang, Jun and others},
  journal={arXiv preprint arXiv:2502.13923},
  year={2025}
}

@inproceedings{gao2022kuairand,
  title={Kuairand: An unbiased sequential recommendation dataset with randomly exposed videos},
  author={Gao, Chongming and Li, Shijun and Zhang, Yuan and Chen, Jiawei and Li, Biao and Lei, Wenqiang and Jiang, Peng and He, Xiangnan},
  booktitle={Proceedings of the 31st ACM international conference on information \& knowledge management},
  pages={3953--3957},
  year={2022}
}

@article{bennett2007netflix,
  title={The netflix prize},
  author={Bennett, James and Lanning, Stan},
  year={2007},
  publisher={Netflix}
    
}

@inproceedings{wu2020mind,
  title={Mind: A large-scale dataset for news recommendation},
  author={Wu, Fangzhao and Qiao, Ying and Chen, Jiun-Hung and Wu, Chuhan and Qi, Tao and Lian, Jianxun and Liu, Danyang and Xie, Xing and Gao, Jianfeng and Wu, Winnie and others},
  booktitle={Proceedings of the 58th annual meeting of the association for computational linguistics},
  pages={3597--3606},
  year={2020}
}

@article{hou2024amazon,
  title={Bridging language and items for retrieval and recommendation},
  author={Hou, Yupeng and Li, Jiacheng and He, Zhankui and Yan, An and Chen, Xiusi and McAuley, Julian},
  journal={arXiv preprint arXiv:2403.03952},
  year={2024}
}

@article{asghar2016yelp,
  title={Yelp dataset challenge: Review rating prediction},
  author={Asghar, Nabiha},
  journal={arXiv preprint arXiv:1605.05362},
  year={2016}
}

@article{ni2023content,
  title={A content-driven micro-video recommendation dataset at scale},
  author={Ni, Yongxin and Cheng, Yu and Liu, Xiangyan and Fu, Junchen and Li, Youhua and He, Xiangnan and Zhang, Yongfeng and Yuan, Fajie},
  journal={arXiv preprint arXiv:2309.15379},
  year={2023}
}

@article{yang2024oasis,
  title={Oasis: Open agent social interaction simulations with one million agents},
  author={Yang, Ziyi and Zhang, Zaibin and Zheng, Zirui and Jiang, Yuxian and Gan, Ziyue and Wang, Zhiyu and Ling, Zijian and Chen, Jinsong and Ma, Martz and Dong, Bowen and others},
  journal={arXiv preprint arXiv:2411.11581},
  year={2024}
}

@inproceedings{zhao2018recommendations,
  title={Recommendations with negative feedback via pairwise deep reinforcement learning},
  author={Zhao, Xiangyu and Zhang, Liang and Ding, Zhuoye and Xia, Long and Tang, Jiliang and Yin, Dawei},
  booktitle={Proceedings of the 24th ACM SIGKDD international conference on knowledge discovery \& data mining},
  pages={1040--1048},
  year={2018}
}

@article{yang2025reccot,
  title={RecCoT: Enhancing Recommendation via Chain-of-Thought},
  author={Yang, Shuo and Cao, Jiangxia and Li, Haipeng and Mao, Yuqi and Pang, Shuchao},
  journal={arXiv preprint arXiv:2506.21032},
  year={2025}
}

@article{han2025reinforcement,
  title={Reinforcement Learning from User Feedback},
  author={Han, Eric and Chen, Jun and Sankararaman, Karthik Abinav and Peng, Xiaoliang and Xu, Tengyu and Helenowski, Eryk and Peng, Kaiyan and Kumar, Mrinal and Wang, Sinong and Fang, Han and others},
  journal={arXiv preprint arXiv:2505.14946},
  year={2025}
}

@article{doubao,
  title={Exploring data scaling trends and effects in reinforcement learning from human feedback},
  author={Shen, Wei and Liu, Guanlin and Wu, Zheng and Zhu, Ruofei and Yang, Qingping and Xin, Chao and Yue, Yu and Yan, Lin},
  journal={arXiv preprint arXiv:2503.22230},
  year={2025}
}

@inproceedings{lai2025dar,
  title={DAR: Dimension-Adaptive Recommendation with Multi-Granular Noise Control},
  author={Lai, Riwei and Chen, Li and Chen, Rui and Zhang, Chi},
  booktitle={Proceedings of the 48th International ACM SIGIR Conference on Research and Development in Information Retrieval},
  pages={2203--2212},
  year={2025}
}

@inproceedings{lai2024adaptive,
  title={Adaptive hardness negative sampling for collaborative filtering},
  author={Lai, Riwei and Chen, Rui and Han, Qilong and Zhang, Chi and Chen, Li},
  booktitle={Proceedings of the AAAI Conference on Artificial Intelligence},
  volume={38},
  number={8},
  pages={8645--8652},
  year={2024}
}

@inproceedings{lai2023disentangled,
  title={Disentangled negative sampling for collaborative filtering},
  author={Lai, Riwei and Chen, Li and Zhao, Yuhan and Chen, Rui and Han, Qilong},
  booktitle={Proceedings of the Sixteenth ACM International Conference on Web Search and Data Mining},
  pages={96--104},
  year={2023}
}

@article{lai2024survey,
  title={A survey on data-centric recommender systems},
  author={Lai, Riwei and Chen, Rui and Zhang, Chi},
  journal={arXiv preprint arXiv:2401.17878},
  year={2024}
}

@article{yang2025qwen3,
  title={Qwen3 technical report},
  author={Yang, An and Li, Anfeng and Yang, Baosong and Zhang, Beichen and Hui, Binyuan and Zheng, Bo and Yu, Bowen and Gao, Chang and Huang, Chengen and Lv, Chenxu and others},
  journal={arXiv preprint arXiv:2505.09388},
  year={2025}
}

@article{sen,
  title={Super Encoding Network: Recursive Association of Multi-Modal Encoders for Video Understanding},
  author={Chen, Boyu and Chen, Siran and Li, Kunchang and Xu, Qinglin and Qiao, Yu and Wang, Yali},
  journal={arXiv preprint arXiv:2506.07576},
  year={2025}
}

@article{pcaby,
  title={Percept, chat, and then adapt: Multimodal knowledge transfer of foundation models for open-world video recognition},
  author={Chen, Boyu and Chen, Siran and Li, Kunchang and Xu, Qinglin and Qiao, Yu and Wang, Yali},
  journal={arXiv preprint arXiv:2402.18951},
  year={2024}
}

@article{videochata1,
  title={VideoChat-A1: Thinking with Long Videos by Chain-of-Shot Reasoning},
  author={Wang, Zikang and Chen, Boyu and Yue, Zhengrong and Wang, Yi and Qiao, Yu and Wang, Limin and Wang, Yali},
  journal={arXiv preprint arXiv:2506.06097},
  year={2025}
}

@article{yue2025uniflow,
  title={UniFlow: A Unified Pixel Flow Tokenizer for Visual Understanding and Generation},
  author={Yue, Zhengrong and Zhang, Haiyu and Zeng, Xiangyu and Chen, Boyu and Wang, Chenting and Zhuang, Shaobin and Dong, Lu and Du, KunPeng and Wang, Yi and Wang, Limin and others},
  journal={arXiv preprint arXiv:2510.10575},
  year={2025}
}

@article{vragent,
  title={VRAgent-R1: Boosting Video Recommendation with MLLM-based Agents via Reinforcement Learning},
  author={Chen, Siran and Chen, Boyu and Yu, Chenyun and Luo, Yuxiao and Yi, Ouyang and Cheng, Lei and Zhuo, Chengxiang and Li, Zang and Wang, Yali},
  journal={arXiv preprint arXiv:2507.02626},
  year={2025}
}

@article{lvagent,
  title={Lvagent: Long video understanding by multi-round dynamical collaboration of mllm agents},
  author={Chen, Boyu and Yue, Zhengrong and Chen, Siran and Wang, Zikang and Liu, Yang and Li, Peng and Wang, Yali},
  journal={arXiv preprint arXiv:2503.10200},
  year={2025}
}

@article{gpt4o,
  title={Gpt-4o system card},
  author={Hurst, Aaron and Lerer, Adam and Goucher, Adam P and Perelman, Adam and Ramesh, Aditya and Clark, Aidan and Ostrow, AJ and Welihinda, Akila and Hayes, Alan and Radford, Alec and others},
  journal={arXiv preprint arXiv:2410.21276},
  year={2024}
}

@article{guo2017deepfm,
  title={DeepFM: a factorization-machine based neural network for CTR prediction},
  author={Guo, Huifeng and Tang, Ruiming and Ye, Yunming and Li, Zhenguo and He, Xiuqiang},
  journal={arXiv preprint arXiv:1703.04247},
  year={2017}
}

@inproceedings{kang2018self,
  title={Self-attentive sequential recommendation},
  author={Kang, Wang-Cheng and McAuley, Julian},
  booktitle={2018 IEEE international conference on data mining (ICDM)},
  pages={197--206},
  year={2018},
  organization={IEEE}
}

@article{yang2024qwen2,
  title={Qwen2. 5 technical report},
  author={Yang, An and Yang, Baosong and Zhang, Beichen and Hui, Binyuan and Zheng, Bo and Yu, Bowen and Li, Chengyuan and Liu, Dayiheng and Huang, Fei and Wei, Haoran and others},
  journal={arXiv preprint arXiv:2412.15115},
  year={2024}
}

@inproceedings{zhang2024agent4rec,
  title={On generative agents in recommendation},
  author={Zhang, An and Chen, Yuxin and Sheng, Leheng and Wang, Xiang and Chua, Tat-Seng},
  booktitle={Proceedings of the 47th international ACM SIGIR conference on research and development in Information Retrieval},
  pages={1807--1817},
  year={2024}
}

@article{corecco2024suber,
  title={SUBER: An RL Environment with Simulated Human Behavior for Recommender Systems},
  author={Corecco, Nathan and Piatti, Giorgio and Lanzend{\"o}rfer, Luca A and Fan, Flint Xiaofeng and Wattenhofer, Roger},
  journal={arXiv preprint arXiv:2406.01631},
  year={2024}
}

@article{guo2025deepseek,
  title={Deepseek-r1: Incentivizing reasoning capability in llms via reinforcement learning},
  author={Guo, Daya and Yang, Dejian and Zhang, Haowei and Song, Junxiao and Zhang, Ruoyu and Xu, Runxin and Zhu, Qihao and Ma, Shirong and Wang, Peiyi and Bi, Xiao and others},
  journal={arXiv preprint arXiv:2501.12948},
  year={2025}
}

@article{wang2023rethinking,
  title={Rethinking the evaluation for conversational recommendation in the era of large language models},
  author={Wang, Xiaolei and Tang, Xinyu and Zhao, Wayne Xin and Wang, Jingyuan and Wen, Ji-Rong},
  journal={arXiv preprint arXiv:2305.13112},
  year={2023}
}

@inproceedings{xiang2024simuser,
  title={SimUser: Generating Usability Feedback by Simulating Various Users Interacting with Mobile Applications},
  author={Xiang, Wei and Zhu, Hanfei and Lou, Suqi and Chen, Xinli and Pan, Zhenghua and Jin, Yuping and Chen, Shi and Sun, Lingyun},
  booktitle={Proceedings of the 2024 CHI Conference on Human Factors in Computing Systems},
  pages={1--17},
  year={2024}
}

@inproceedings{he2016vista,
  title={Vista: A visually, socially, and temporally-aware model for artistic recommendation},
  author={He, Ruining and Fang, Chen and Wang, Zhaowen and McAuley, Julian},
  booktitle={Proceedings of the 10th ACM conference on recommender systems},
  pages={309--316},
  year={2016}
}

@article{harper2015movielens,
  title={The movielens datasets: History and context},
  author={Harper, F Maxwell and Konstan, Joseph A},
  journal={Acm transactions on interactive intelligent systems (tiis)},
  volume={5},
  number={4},
  pages={1--19},
  year={2015},
  publisher={Acm New York, NY, USA}
}

@article{wang2025user,
  title={User behavior simulation with large language model-based agents},
  author={Wang, Lei and Zhang, Jingsen and Yang, Hao and Chen, Zhi-Yuan and Tang, Jiakai and Zhang, Zeyu and Chen, Xu and Lin, Yankai and Sun, Hao and Song, Ruihua and others},
  journal={ACM Transactions on Information Systems},
  volume={43},
  number={2},
  pages={1--37},
  year={2025},
  publisher={ACM New York, NY}
}

@article{video-r1,
  title={Video-r1: Reinforcing video reasoning in mllms},
  author={Feng, Kaituo and Gong, Kaixiong and Li, Bohao and Guo, Zonghao and Wang, Yibing and Peng, Tianshuo and Wu, Junfei and Zhang, Xiaoying and Wang, Benyou and Yue, Xiangyu},
  journal={arXiv preprint arXiv:2503.21776},
  year={2025}
}

@article{li2025videochat-r1,
  title={Videochat-r1: Enhancing spatio-temporal perception via reinforcement fine-tuning},
  author={Li, Xinhao and Yan, Ziang and Meng, Desen and Dong, Lu and Zeng, Xiangyu and He, Yinan and Wang, Yali and Qiao, Yu and Wang, Yi and Wang, Limin},
  journal={arXiv preprint arXiv:2504.06958},
  year={2025}
}

@article{zhao2025reason,
  title={Reason-to-Recommend: Using Interaction-of-Thought Reasoning to Enhance LLM Recommendation},
  author={Zhao, Keyu and Xu, Fengli and Li, Yong},
  journal={arXiv preprint arXiv:2506.05069},
  year={2025}
}

@inproceedings{park2022exploiting,
  title={Exploiting negative preference in content-based music recommendation with contrastive learning},
  author={Park, Minju and Lee, Kyogu},
  booktitle={Proceedings of the 16th ACM Conference on Recommender Systems},
  pages={229--236},
  year={2022}
}

@article{paudel2018loss,
  title={Loss aversion in recommender systems: Utilizing negative user preference to improve recommendation quality},
  author={Paudel, Bibek and Luck, Sandro and Bernstein, Abraham},
  journal={arXiv preprint arXiv:1812.11422},
  year={2018}
}

@inproceedings{xie2021deep,
  title={Deep feedback network for recommendation},
  author={Xie, Ruobing and Ling, Cheng and Wang, Yalong and Wang, Rui and Xia, Feng and Lin, Leyu},
  booktitle={Proceedings of the twenty-ninth international conference on international joint conferences on artificial intelligence},
  pages={2519--2525},
  year={2021}
}

@article{chen2021curriculum,
  title={Curriculum disentangled recommendation with noisy multi-feedback},
  author={Chen, Hong and Chen, Yudong and Wang, Xin and Xie, Ruobing and Wang, Rui and Xia, Feng and Zhu, Wenwu},
  journal={Advances in Neural Information Processing Systems},
  volume={34},
  pages={26924--26936},
  year={2021}
}

@inproceedings{zhang2024generative,
  title={On generative agents in recommendation},
  author={Zhang, An and Chen, Yuxin and Sheng, Leheng and Wang, Xiang and Chua, Tat-Seng},
  booktitle={Proceedings of the 47th international ACM SIGIR conference on research and development in Information Retrieval},
  pages={1807--1817},
  year={2024}
}

@article{zhao2023kuaisim,
  title={KuaiSim: A comprehensive simulator for recommender systems},
  author={Zhao, Kesen and Liu, Shuchang and Cai, Qingpeng and Zhao, Xiangyu and Liu, Ziru and Zheng, Dong and Jiang, Peng and Gai, Kun},
  journal={Advances in Neural Information Processing Systems},
  volume={36},
  pages={44880--44897},
  year={2023}
}

@inproceedings{zhang2024usimagent,
  title={Usimagent: Large language models for simulating search users},
  author={Zhang, Erhan and Wang, Xingzhu and Gong, Peiyuan and Lin, Yankai and Mao, Jiaxin},
  booktitle={Proceedings of the 47th International ACM SIGIR Conference on Research and Development in Information Retrieval},
  pages={2687--2692},
  year={2024}
}

@article{piao2025agentsociety,
  title={Agentsociety: Large-scale simulation of llm-driven generative agents advances understanding of human behaviors and society},
  author={Piao, Jinghua and Yan, Yuwei and Zhang, Jun and Li, Nian and Yan, Junbo and Lan, Xiaochong and Lu, Zhihong and Zheng, Zhiheng and Wang, Jing Yi and Zhou, Di and others},
  journal={arXiv preprint arXiv:2502.08691},
  year={2025}
}

@article{liu2025mosaic,
  title={Mosaic: Modeling social ai for content dissemination and regulation in multi-agent simulations},
  author={Liu, Genglin and Le, Vivian and Rahman, Salman and Kreiss, Elisa and Ghassemi, Marzyeh and Gabriel, Saadia},
  journal={arXiv preprint arXiv:2504.07830},
  year={2025}
}
\newpage

\section{Appendix of the Paper}
This is the appendix of paper G-UBS: Towards Robust Understanding of Implicit Feedback via Group-Aware User Behavior Simulation.

\subsection{Dataset filtering Method.} To mitigate noise in implicit feedback, such as low completion rates caused by interruptions (e.g., phone calls, driving) or short viewing durations unrelated to user disliking, We filtered the data to retain only implicit negative feedback cases meeting three criteria: (i) user operations both before and after feedback; (ii) an interval within 2 minutes between these operations; and (iii) no exit from the Tencent Video APP.
In this way, we retained the fast-skipping data unrelated to external force majeure factors. 
We also exclude videos with fewer than 10 views (due to a large proportion of poor quality or short platform retention time), resulting in a final set of 25K videos.

\subsection{Multi-modal ablation of G-UBS}

To verify the effectiveness of visual information in UFM, we conducted an ablation experiment without providing visual information, as shown in Tab~\ref{visual} in the Appendix. Under the same training settings, the scenario without visual information achieved suboptimal results, thus demonstrating the importance of multimodal information.
\noindent\textbf{Different Reward Weighting on UFM.}
We examine the effects of varying the individual reward weight $W_T$ from 0.5 to 0.8. As shown in Table~\ref{weight_ab}, the combination of $W_T$=0.7, $W_G$=0.15, and $W_S$=0.15 yields the optimal performance.
Moderate values of $W_G$ and $W_S$ create a meaningful influence on the results, highlighting the critical role of group profiles. Excessively high $W_G$ and $W_S$ lead to the neglect of individual differences, while insufficient values fail to effectively filter noise from users' implicit feedback, leading to suboptimal outcomes.

\subsection{Ablation on the Reward in UFM}

To validate the importance of rewards, we conducted ablation experiments as shown in Tab~\ref{reward_ablation} in the Appendix. Without the reason reward $r_{reason}$, the model’s reasoning capability would be significantly degraded, resulting in poor performance, so the corresponding results are not presented. 
Removing the format reward $r_{format}$ leads to almost no change in performance.
However, it disrupts the expected output format, complicating downstream result parsing and analysis.
Omitting the skip reward $r_{skip}$ results in a noticeable performance drop, confirming its effectiveness.

\subsection{Ablation on the Reflection Process on UGM.}
We remove the `reflect' mechanism in UGM and provide results
in Tab~\ref{reflect}. we achieved significant performance improvement, as this reflection process improves group consistency.

\subsection{Deployment Details and Efficiency}
To improve response speed, we deployed G-UBS with FP16 quantization. Specifically, UGM can analyze 1000 users within 8 minutes through the asynchronous invocation mechanism; meanwhile, supported by 4 A100-80G GPUs, UFM achieves a Queries Per Second of 5.3. As a result, this service can process approximately 458K videos per day.

\begin{table}[!]

    \small
    \setlength{\tabcolsep}{1.8mm}
    \renewcommand{\arraystretch}{0.7}
    \centering
    \begin{tabular}{c|ccccc}
    \toprule
    Reflection & \begin{tabular}[c]{@{}c@{}}Person\\Play Rate  \end{tabular}    & \begin{tabular}[c]{@{}c@{}}Play Rate\\ \textgreater 30\% \end{tabular} & \begin{tabular}[c]{@{}c@{}}Click\\Rate \end{tabular}  & \begin{tabular}[c]{@{}c@{}}Judge\\ F1  \end{tabular}  & \begin{tabular}[c]{@{}c@{}}Reason\\ F1  \end{tabular} \\ 
    \midrule
     \XSolidBrush &   51.2\%   &  87.9\%  & 25.2\% &   54.4\%  &  55.1\%\\ 
     \rowcolor{gray!20}
    \Checkmark &   \textbf{52.3\%}  &  \textbf{88.7\%} &  \textbf{25.7\%}    &  \textbf{54.9\%}   &  \textbf{55.6\%}  \\ \bottomrule
    \end{tabular}
    \caption{Ablation on Reflection of UGM}
    \label{reflect}
\end{table}

\begin{table}[!]

    \small
    \setlength{\tabcolsep}{1.8mm}
    \renewcommand{\arraystretch}{0.7}
    \centering
    \begin{tabular}{c|ccccc}
    \toprule
    Vision info & \begin{tabular}[c]{@{}c@{}}Person\\Play Rate  \end{tabular}    & \begin{tabular}[c]{@{}c@{}}Play Rate\\ \textgreater 30\% \end{tabular} & \begin{tabular}[c]{@{}c@{}}Click\\Rate \end{tabular}  & \begin{tabular}[c]{@{}c@{}}Judge\\ F1  \end{tabular}  & \begin{tabular}[c]{@{}c@{}}Reason\\ F1  \end{tabular} \\ 
    \midrule
     \XSolidBrush &   50.9\%   &  87.4\%  & 24.7\% &   53.5\%  &  54.1\%\\ 
     \rowcolor{gray!20}
    \Checkmark &   \textbf{52.3\%}  &  \textbf{88.7\%} &  \textbf{25.7\%}    &  \textbf{54.9\%}   &  \textbf{55.6\%}  \\ \bottomrule
    \end{tabular}
    \caption{Ablation on Multi-modal Information}
    \label{visual}
\end{table}

\begin{table}[!]
    \small
    \setlength{\tabcolsep}{0.7mm}
    \renewcommand{\arraystretch}{0.7}
    \centering
    \begin{tabular}{ccc|ccccc}
    \toprule
   \begin{tabular}[c]{@{}c@{}}Format\\Reward  \end{tabular}  & \begin{tabular}[c]{@{}c@{}}Skip\\Reward  \end{tabular} & \begin{tabular}[c]{@{}c@{}}Reason\\Reward  \end{tabular} & \begin{tabular}[c]{@{}c@{}}Person\\Play Rate  \end{tabular}    & \begin{tabular}[c]{@{}c@{}}Play Rate\\ \textgreater 30\% \end{tabular} & \begin{tabular}[c]{@{}c@{}}Click\\Rate \end{tabular}  & \begin{tabular}[c]{@{}c@{}}Judge\\ F1  \end{tabular}  & \begin{tabular}[c]{@{}c@{}}Reason\\ F1  \end{tabular} \\ 
    \midrule
    \XSolidBrush & \XSolidBrush & \Checkmark & 51.9\%   &  88.4\%  & 25.0\% &   54.3\%  &  55.2\%\\ 
     \Checkmark & \XSolidBrush & \Checkmark & 51.9\%   &  88.5\%  & 25.2\% &   54.3\%  &  55.2\%\\ 
     \XSolidBrush & \Checkmark & \Checkmark & 52.2\%   &  88.6\%  & 25.5\% &   54.8\%  &  55.3\%\\ 
     \rowcolor{gray!20}
    \Checkmark & \Checkmark & \Checkmark & \textbf{52.3\%}  &  \textbf{88.7\%} &  \textbf{25.7\%}    &  \textbf{54.9\%}   &  \textbf{55.6\%}  \\ \bottomrule
    \end{tabular}
    \caption{Ablation on the Reward in UFM.}
    \label{reward_ablation}
\end{table}

\begin{table}[!]
    \small
    \setlength{\tabcolsep}{1.1mm}
    \renewcommand{\arraystretch}{1}
    \begin{tabular}{ccc|ccccc}
    \toprule
    $W_T$ & $W_G$ & $W_S$& \begin{tabular}[c]{@{}c@{}}Person\\Play Rate  \end{tabular}    & \begin{tabular}[c]{@{}c@{}}Play Rate\\ \textgreater 30\% \end{tabular} & \begin{tabular}[c]{@{}c@{}}Click\\Rate \end{tabular}  & \begin{tabular}[c]{@{}c@{}}Judge\\ F1  \end{tabular}  & \begin{tabular}[c]{@{}c@{}}Reason\\ F1  \end{tabular}\\ 
    \midrule
     0.5 & 0.25 &  0.25 & 51.6\% & 88.4\% & 25.2\%  & 53.8\% & 54.8\%\\ 
     0.6 & 0.2 &  0.2 & 51.8\% & 88.2\% & 25.1\%  & 54.0\% & 55.0\%\\ 
     \rowcolor{gray!20}
    \textbf{0.7} & \textbf{0.15} & \textbf{0.15} &  \textbf{52.3\%} & \textbf{88.7\%} & \textbf{25.7\%}  & \textbf{54.9\%} & \textbf{55.6\%}  \\ 
    0.8 & 0.1 & 0.1 &  52.1\% & 88.7\% & 25.1\%  & 54.2\% & 54.9\%\\ \bottomrule
    \end{tabular}
    \caption{Ablation on Reward Weighting on UFM}
    \label{weight_ab}
\end{table}

\newpage
\subsection{A.1 Prompt for UGM Agents}
1. Prompt for group ``summarize" process in UGM. The summary mode specifies the grouping criteria, such as video preferences
or demographic attributes (e.g., age, gender, occupation). 
All user profiles is a user profile set $\mathcal{U}$ containing over 1000 users, each user is formatted as $u = [\text{ID}, \text{Occ}, \text{Age}, \text{Gender}, \mathbf{T}],\space u \in \mathcal{U}$.
Here, $\text{Occ}$ denotes the user's occupation and $\mathbf{T}$ represents tags for the user's video preferences. All user profiles are passed into the prompt in the form of a list.

\noindent\fbox{\parbox{\linewidth}{
\textbf{User prompt:}

Given a group that includes the users' age, gender, occupation, video interest tags, and other information (e.g., marital status, city size, etc).

Please divide this group into \textbf{\{group number\}} categories. 

For classification, conduct a comprehensive analysis based on \textbf{\{summary mode\}}
Please clarify the criteria for classification, so that users with similar characteristics can be aggregated into the same category. 
Please output the definition and characteristics of each category (with characteristics as detailed as possible), and who the typical user of the category is.

Input information \textbf{\{all user profiles\}}

Please strictly follow the output format and do not add bold formatting, enumeration commas, or other special symbols in the output.
Output format must comply with the following:

\textbf{User definitions and typical features:}

Category 1: xxx

Features: xxx

Typical user: User xxx

Category 2: xxx

Features: xxx

Typical user: User xxx

Category 3: xxx

Features: xxx

Typical user: User xxx

}}
\newpage

2. Prompt for profile generator in the ``cluster" and the ``reflect" process in UGM. Group name and group characteristics are generated from the ``summarize" process. Group user profiles are the profiles in the group. 
Taking group $C_g$ as an example: for each user, their profiles $u = [\text{ID}, \text{Occ}, \text{Age}, \text{Gender}, \mathbf{T}],\space u \in C_g$
In the ``cluster" process, we do not input the viewing histories in group user profiles; in the ``reflect" process, we will input all the users' histories (e.g., play rate, video title, duration, click). 

\noindent\fbox{\parbox{\linewidth}{
\textbf{User prompt:}

Given the user profile in a group of users and the group characteristics, please generate the overall modification suggestions for this group.

The name of this group is: \textbf{\{group name\}}

The characteristics of this group are:\textbf{ \{group characteristics\}}.

Please, based on the characteristics of this group, provide the overall description and modification suggestions for the group. Clearly explain how to improve the system’s recommendation performance for this group.

You can focus on how to refine the tags of users' preferences, what types of video content this group is not interested in, and what types of video content they are interested in.
All information about the user in this group is as follows: \textbf{\{group user profiles \}}

}}

3. Prompt for judging whether the user matches the group profiles in the ``reflect" process in UGM. 

\noindent\fbox{\parbox{\linewidth}{
\textbf{User prompt:}

Given a person's user profile, including gender, viewing history, interest tags, and the recommendation modification suggestions for this person. This person belongs to the \textbf{\{group name\}} group. Determine whether the group profile and modification suggestions for this group can be applied to this person.

Please strictly follow the output format below. Do not add any special symbols, enumeration commas, or bold formatting.

Age: \textbf{\{user age\}}

Gender: \textbf{\{user gender\}}

Occupation: \textbf{\{user occupation\}}

Other information \textbf{\{user other information\}}

User’s video interest tags: \textbf{\{interest tags\}}

history: \textbf{\{User history\}}

Group’s profile: \textbf{\{group profile\}}

\textbf{Output format:}

Match or not: Yes/No

Reason: xxx

}}

\newpage

\subsection{A.2 Prompt for UFM Agents}
Prompt for the generation of $O_T$ (the output corresponding to the training user profile) and $O_S$ (the output corresponding to a similar user profile) in UFM shown in Fig 3 in the main text. The prompt for inference also uses this prompt. The video input consists of sampled frames from the watched video, while other user information includes details such as marital status (single, married, or parenting), place of residence (small towns or cities), and occupation type (blue-collar or white-collar), among others.

\noindent\fbox{\parbox{\linewidth}{
\textbf{User prompt for sequential recommendation task in IF-VR:}

This is a User profile: 

Age: \textbf{\{user age\}}

Gender: \textbf{\{user gender\}}

Occupation: \textbf{\{user occupation\}}

Other information \textbf{\{user other information\}}

User’s video interest tags: \textbf{\{interest tags\}}

history: \textbf{\{user history\}}

This is the information of the currently watched video:

Title: \textbf{\{watched video title\}}

Video duration: \textbf{\{video length\}}

Video vision info \textbf{\{video input\}}

\textbf{Based on the user profile, viewing histories, and current video, judge if the user will quickly skip this video. }

\textbf{We provide 4 options: }

\textbf{A:} The user will not fast-skip this video.

\textbf{B:} Content-driven fast-skipping (due to objective content flaws) e.g., vulgar content, clickbait titles, physiologically discomforting visuals (e.g., bloody scenes, unpleasant creatures like snakes or centipedes).

\textbf{C: }Algorithm-driven fast-skipping (due to recommendation ineffectiveness) e.g., inaccurate user profiling, repetitive recommendations, insufficient diversity.

\textbf{D:} User-driven fast-skipping (arising from individual user actions)
such as operational errors and a lack of viewing intent at the current time.

You can provide the analysis from the aspects of values, interest matching, tolerance for negative energy, algorithm matching, content quality, and the user's own reasons. 

Please think about this question as if you were a human pondering deeply.

It's encouraged to include self-reflection or verification in the reasoning process.

Provide your reasoning between the  \textless think\textgreater  \textless /think\textgreater tags, and then give your final answer between the  \textless answer\textgreater  \textless /answer\textgreater tags.

}}

\newpage

\noindent\fbox{\parbox{\linewidth}{
\textbf{User prompt for click simulation task in IF-VR:}

This is a User profile: 

Age: \textbf{\{user age\}}

Gender: \textbf{\{user gender\}}

Occupation: \textbf{\{user occupation\}}

Other information \textbf{\{user other information\}}

User’s video interest tags: \textbf{\{interest tags\}}

history: \textbf{\{user history\}}

This is the information of the currently watched video:

Title: \textbf{\{watched video title\}}

Video duration: \textbf{\{video length\}}

Video vision info \textbf{\{video's first frame\}}

\textbf{Based on the user profile, viewing histories, and current video, judge if the user will click this video. }

\textbf{We provide 2 options: }

\textbf{A:} The user will click this video.

\textbf{B:} The user will not click this video.

You can provide the analysis from the aspects of values, interest matching, tolerance for negative energy, algorithm matching, content quality, and the user's own reasons. 

Please think about this question as if you were a human pondering deeply.

It's encouraged to include self-reflection or verification in the reasoning process.

Provide your reasoning between the  \textless think\textgreater  \textless /think\textgreater tags, and then give your final answer between the  \textless answer\textgreater  \textless /answer\textgreater tags.

}}

\newpage

Prompt for the generation of $O_G$ in the sequential recommendation task (the output corresponding to the group profile) in Fig 3 in the main text. 

\noindent\fbox{\parbox{\linewidth}{
\textbf{User prompt:}

This is a group profile: 

\textbf{\{group name\}}, \textbf{\{group profile\}}: 

Title of the currently watched video: \textbf{\{watched video title\}}
Video duration: \textbf{\{video length\}}

\textbf{Based on the group profile, viewing histories, and current video, judge if the user in this group will quickly skip this video. }

\textbf{We provide 4 options: }

\textbf{A:} The user will not fast-skip this video.

\textbf{B:} Content-driven fast-skipping (due to objective content flaws) e.g., vulgar content, clickbait titles, physiologically discomforting visuals (e.g., bloody scenes, unpleasant creatures like snakes or centipedes).

\textbf{C: }Algorithm-driven fast-skipping (due to recommendation ineffectiveness) e.g., inaccurate user profiling, repetitive recommendations, insufficient diversity.

\textbf{D:} User-driven fast-skipping (arising from individual user actions)
such as operational errors and a lack of viewing intent at the current time.

You can provide the analysis from the aspects of values, interest matching, tolerance for negative energy, algorithm matching, and content quality. 

Please think about this question as if you were a human pondering deeply.

It's encouraged to include self-reflection or verification in the reasoning process.

Provide your reasoning between the  \textless think\textgreater  \textless /think\textgreater tags, and then give your final answer between the  \textless answer\textgreater  \textless /answer\textgreater tags.

}}

Prompt for the generation of $O_G$ in click simulation task (the output corresponding to the group profile) in Fig 3 in the main text. 

\noindent\fbox{\parbox{\linewidth}{
\textbf{User prompt:}

This is a group profile: 

\textbf{\{group name\}}, \textbf{\{group profile\}}: 

Title of the currently watched video: \textbf{\{watched video title\}}
Video duration: \textbf{\{video length\}}

\textbf{Based on the group profile, viewing histories, and current video, judge if the user in this group will click this video. }

\textbf{We provide 2 options: }

\textbf{A:} The user will click this video.

\textbf{B:} The user will not click this video.

You can provide the analysis from the aspects of values, interest matching, tolerance for negative energy, algorithm matching, and content quality. 

Please think about this question as if you were a human pondering deeply.

It's encouraged to include self-reflection or verification in the reasoning process.

Provide your reasoning between the  \textless think\textgreater  \textless /think\textgreater tags, and then give your final answer between the  \textless answer\textgreater  \textless /answer\textgreater tags.

}}

\begin{figure*}[t]
    \centering
    \includegraphics[width=1\linewidth]{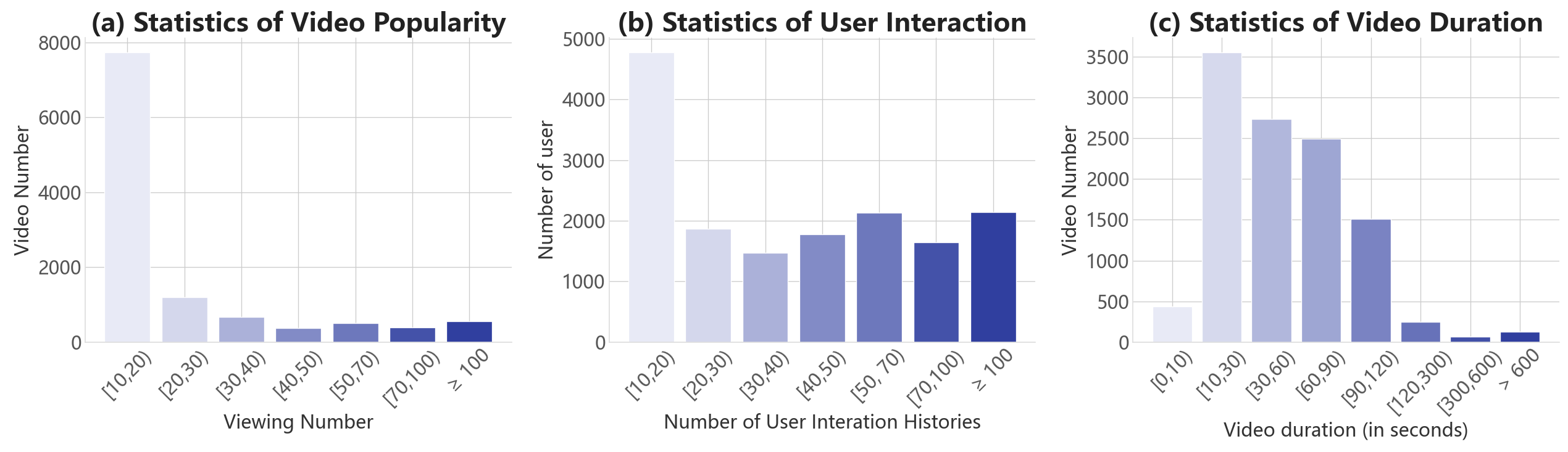}

    \caption{Visualization of IF-VR Dataset.}    
    \label{fig:dataset_vis}
    
\end{figure*}

\subsection{A.3 Additional Details for IF-VR Construction}

\paragraph{SFT Data }We utilized GPT to generate the reasons why a user disliked a video based on their real dislike data, categorizing these reasons into three types: algorithm-driven, video content-driven, or user-driven (e.g., repeatedly recommending game videos to a user seeking learning materials). This annotated data was used as the supervised fine-tuning (SFT) data for the User Feedback Modeler (UFM). shown in the section other prompt.

\paragraph{Sequential Recommendation Task:} For the sequential recommendation task, we selected 1,000 users as the test set, with the remaining users serving as the training set. 
For each user, the first half of their viewing history was used as the input context, and the model was tasked with simulating the user to predict two outcomes for the second half of the history: whether the user would fast-skip the video, and if so, the underlying reason for the fast-skipping. We generated 72K annotations for fast-skipping behaviors. Specifically, GPT-4o was first used to determine the cause of a user’s fast-skipping (algorithm-driven, content-driven, or user-driven) and provide a rationale. Subsequently, GPT-4o was instructed to reflect on the validity of its own reasoning. This process was repeated up to 3 times; if the reasoning remained deemed invalid after 3 rounds of reflection, manual annotation was conducted. Finally, we produce 72K implicit feedback annotations.

\paragraph{Click Simulation Task:} For the click prediction task, similar to SimUser, we only predicted whether a user would click on a video, without analyzing the reasons for non-clicks. This decision was based on the large volume of exposure-click data and its higher noise level—users often scan only a few videos on a page, meaning non-clicks may result from failure to notice the content rather than genuine disinterest. 
We record the video presented on the page recommendation and which video is clicked by user.
We excluded users with fewer than 30 exposure-click records, resulting in 613K interaction records from 7,000 users. Among these, 6,000 users were used for training, and the test set was constructed using the last 6 click records of the remaining 1,000 users.

\subsection{A.4 Visualization of IF-VR}

To better visualize our dataset, we present three-dimensional visualizations of it, as shown in Fig~\ref{fig:dataset_vis}.
Panel (a) in Fig~\ref{fig:dataset_vis} illustrates the visualization of video popularity. It can be observed that in IF-VR, videos with 10–20 view counts among the collected users are the most numerous.
Panel (b) in Fig~\ref{fig:dataset_vis} displays statistics on the number of users' historical interaction records. The largest number of users have 10–20 interaction records.
Panel (c) in Fig~\ref{fig:dataset_vis} presents the length of watched videos. In news scenarios, videos with a duration of 10–30 seconds account for the highest proportion.

\subsection{A.5 Training details for UFM}

The training pipeline is built on the open-rl framework with an efficient codebase, running on 4 NVIDIA A100 80G GPUs. It adopts bf16 precision and integrates Zero2, Flash-Attention, and gradient checkpointing techniques for enhanced efficiency. For inference, the configuration uses a temperature of 1.0, top\_p of 1.0, and top\_k of 1.0, with VLLM employed for acceleration; the maximum generation length is set to 512. During training, the batch size is 4, and each prompt generates 6 samples (1 for the group profile, 1 for the similar user, and 4 for the training user). The Adam optimizer is used with a learning rate of 1e-5, a weight decay of 0.01, and a beta value of 0.04.

\subsection{A.6 Detailed Experiment Setting of IF-VR}

\paragraph{Sequential video recommendation setting}

The experiment designates 1000 users as test users, with the remaining users serving as training users. The setup involves the following key steps. 

\noindent\textbf{Group Profile Generation: }
To generate group profiles, 1000 users are selected from the training data, and group profiles are generated using the UGM Agent. This selection strategy avoids exceeding the maximum token limit of Deepseek~\cite{guo2025deepseek} when processing large numbers of input users. The generated group profiles are used to guide the training of the UFM model. 

\noindent\textbf{Training Setting: }During training, strict control is applied to ensure the number of samples for each of the four options (A, B, C, D) is balanced. This prevents reward hacking, where the model might otherwise prioritize the most frequent option without genuine learning.

\noindent\textbf{Testing Setting: }
Testing involves inputting user profiles and video information into the model. Due to the minimal distribution gap between test and training users, the classification criteria applied to training users are directly extended to test users. Specifically, the first half of each user’s viewing history is used as contextual input, while the second half serves as test data. If a video in the user's history is predicted to be uninteresting by UFM, this viewing history is excluded from calculations of finish rate and play rate. Additionally, the task evaluates Judge F1/Acc and Reason F1/Acc to evaluate the model’s ability to predict skips and identify why users skip a video.

\paragraph{Click Simulation setting}

The experiment designates 1000 users as test users, with the remaining users serving as training users. The setup involves the following key steps:

\noindent\textbf{SFT Data Preparation: }
For this task, the SFT data adopts a binary classification setup with two options: (A) "The user will click this video" and (B) "The user will not click this video". We randomly select 1000 users and use GPT-4o to annotate the reasons behind users' clicks on videos, thereby constructing the SFT data. The specific generation method is similar to that of the SFT Data described in Section A.3.

\noindent\textbf{Group Profile Generation: }
The method for generating group profiles is consistent with that in the Sequential Video Recommendation setting. In both the training and testing phases of the UFM model, the input video information includes the video's cover frame and title.

\noindent\textbf{Testing Setting}
Testing involves inputting user profiles and video information into the model. Due to the minimal distribution gap between test and training users, the classification criteria applied to training users are directly extended to test users. Specifically, the last 10 historical messages of 1000 users serve as the test set, and the rest serve as the users' interaction history. In terms of performance metrics: If a video in the user's historical records is predicted by UFM as one that the user will not click on, this historical record will be excluded from the calculation of the average click rate. Additionally, we calculate the overall click rate specifically for videos that UFM predicts users will click on.

\subsection{A.7 Detailed Experiment Setting on MovieLens and Amazon Books}

\paragraph{User profile preparation:} Specifically, we randomly select 50 items from users with click histories in the Movielens and Amazon Books datasets. Items rated $\geq$4 by users are labeled as "liked", and those rated $\leq$2 as "disliked"; a large language model (LLM) is then used to generate the user's profile. Among these, items with user ratings $\geq$4 and click records are defined as "interested", while items with low ratings or no click records are defined as "uninterested".

\paragraph{Group profiles preparation}: With these user profiles, we use 1000 users from the training set and leverage the UGM agent to generate group profiles and typical users. 

\paragraph{Training settings:} Following the above operations, the UFM agent can be used for RL training. During training, we conduct reinforcement learning (RL) based on the SFT model from the IF-VR click task. The training task is designed as a binary choice: given an item and a user's profile, the UFM agent is required to determine whether the user has clicked on the item (i.e., whether the user is interested in the item).

\paragraph{Testing settings:} In the testing phase, we adopt a strategy consistent with the SimUser method to verify the UFM agent's ability to identify user-clicked items (i.e., interested items). To avoid label leakage, the test items selected do not appear in the items used for generating user profiles. Specifically, each user is presented with 10 clicked (interested) and 10 unclicked (uninterested) items, and the UFM agent is tasked with distinguishing the clicked (i.e., interested) items. The overall experimental setup is fully aligned with the SimUser~\cite{xiang2024simuser}.

\newpage

\subsection{A.8 Detailed Visualization of UGM Agent}

This section presents the detailed group profiles for health and wellness followers, generated by the UGM Agent. We will release all the group profiles after this article is published. 

\subsubsection{Final Group profile for Health and wellness follower:} 

Based on the characteristics of the health and wellness enthusiast group and user feedback data, we have clarified the preferences and behavioral patterns of this group. The following analysis outlines how to improve the recommendation system to meet the needs of this group.

\textbf{1. Summary of Group Characteristics and Content Preferences}

\textbf{1.1 Core Characteristics of the Group}
\begin{itemize}[leftmargin=*, label={-}]
    \item \textbf{Age and gender}: Primarily female retirees aged 60–70.  
    \item \textbf{Living conditions}: Value physical health and quality of life, with a focus on dietary health, disease prevention, health maintenance, and mental well-being.  
    \item \textbf{Areas of interest}: Health, people’s livelihood news, emotional content, and life tips.  
    \item \textbf{Emotional needs}: Prefer positive emotional content and have low tolerance for negative or conflicting topics.  
\end{itemize}

\textbf{1.2 Content Preferences}
\begin{itemize}[leftmargin=*, label={-}]
    \item \textbf{High-interest areas}:  
    \begin{itemize}[leftmargin=*, label={-}]
        \item Health and wellness content: Disease prevention, exercise, and dietary health care.  
        \item Emotional content: Stories about family and romantic relationships.  
        \item Practical life content: Food tutorials and life hacks.  
        \item Animals and plants: Particularly pet-related content.  
    \end{itemize}

    \item \textbf{Medium-interest areas}:  
    \begin{itemize}[leftmargin=*, label={-}]
        \item Entertainment: Moderate attention to celebrity updates, with a preference for content connected to daily life.  
        \item Social news: Disaster incidents and public welfare events, but presented in a lightweight manner.  
    \end{itemize}

    \item \textbf{Low-interest areas}:  
    \begin{itemize}[leftmargin=*, label={-}]
        \item Overly academic or complex content (e.g., technical in-depth analyses).  
        \item Negative or conflicting content (e.g., social conflicts or tragic events).  
        \item Long videos or those with a slow pace.  
    \end{itemize}
\end{itemize}  

\textbf{2. Promotion Strategies and Recommendation Optimization Goals}

\textbf{2.1 Optimization Goals}
\begin{itemize}[leftmargin=*, label={-}]
    \item \textbf{Improve user completion rate}:  
    \begin{itemize}[leftmargin=*, label={-}]
        \item Reduce mid-video drop-off rates by enhancing content relevance, optimizing video duration, and increasing opening appeal.  
    \end{itemize}

    \item \textbf{Enhance user interaction experience}:  
    \begin{itemize}[leftmargin=*, label={-}]
        \item Deliver high-quality content aligned with individual interests while avoiding negative emotional impacts.  
    \end{itemize}

    \item \textbf{Strengthen personalized recommendations}:  
    \begin{itemize}[leftmargin=*, label={-}]
        \item Dynamically adjust tag weights and integrate refined interest areas into the recommendation algorithm.  
    \end{itemize}
\end{itemize}

\textbf{3. Improvement Directions for the Recommendation System}

\textbf{3.1 Tag Refinement}

\begin{enumerate}[leftmargin=*, label={\arabic*.}]
    \item \textbf{Health and wellness tags} (high weight, priority recommendation):  
    \begin{itemize}[leftmargin=*, label={-}]
        \item \textbf{Dietary health}: Content such as therapeutic diets, nutritional matching, and anti-aging ingredients.  
        \item \textbf{Disease prevention}: Popular science on disease management methods (e.g., prevention of cardiovascular diseases and diabetes).  
        \item \textbf{Mental health}: Elderly psychological counseling and tips for maintaining a positive mindset.  
        \item \textbf{Health maintenance tips}: Simple and practical advice suitable for users (e.g., massage and traditional health practices).  
    \end{itemize}

    \item \textbf{Emotional tags} (medium-high weight):  
    \begin{itemize}[leftmargin=*, label={-}]
        \item \textbf{Family stories}: Positive stories about intergenerational relationships.  
        \item \textbf{Love \& friendship}: Warm emotional videos (e.g., stories of long-lived couples).  
        \item \textbf{Positive energy dissemination}: Public welfare events and touching moments.  
    \end{itemize}

    \item \textbf{People’s livelihood news and life tags} (medium weight):  
    \begin{itemize}[leftmargin=*, label={-}]
        \item \textbf{People’s livelihood issues}: Low-conflict updates on people’s livelihood, presented in a lightweight manner.  
        \item \textbf{Life hacks}: Highly practical knowledge (e.g., money-saving tips and housework optimization).  
        \item \textbf{Food tutorials}: Simple-to-operate home recipes and recommendations for health-preserving ingredients.  
    \end{itemize}

    \item \textbf{Technology and animal/plant tags} (secondary weight):  
    \begin{itemize}[leftmargin=*, label={-}]
        \item \textbf{Science popularization}: Content on technology applications relevant to the daily lives of elderly users (e.g., tips for using smart devices).  
        \item \textbf{Pet interaction}: Priority to warm and cute pet stories.  
    \end{itemize}

    \item \textbf{Tags with reduced weight}:  
    \begin{itemize}[leftmargin=*, label={-}]
        \item In-depth academic analyses (e.g., complex social analyses or technical demonstrations).  
        \item Negative emotional news (including disasters and conflicts).  
        \item Long videos or those with overly provocative titles (such titles may attract attention but reduce completion rates).  
    \end{itemize}
\end{enumerate}

\textbf{3.2 Duration Optimization}

To align with the group’s behavioral habits, the recommendation system should filter videos by duration:  
\begin{itemize}[leftmargin=*, label={-}]
    \item \textbf{Optimal duration range}:  
    \begin{itemize}[leftmargin=*, label={-}]
        \item 15–60 seconds: Concise videos are more likely to achieve high completion rates.  
        \item Long videos (\textgreater90 seconds): Need to be content-dense and get to the point directly.  
    \end{itemize}

    \item \textbf{Enhance first-screen appeal}:  
    \begin{itemize}[leftmargin=*, label={-}]
        \item Key information should be highlighted within the first 5 seconds to avoid lengthy introductions.  
    \end{itemize}
\end{itemize}  

\textbf{3.3 Negative Emotion Management}

Through emotional analysis, negative content should be filtered:  
\begin{itemize}[leftmargin=*, label={-}]
    \item Exclude negative events with strong emotional impact.  
    \item Disseminate disaster news in a lightweight manner, emphasizing solutions or positive aspects.  
    \item Reduce the recommendation frequency of socially controversial topics.  
\end{itemize}  

\textbf{3.4 Personalized Dynamic Adjustment}

Implement dynamic personalized adjustment strategies:  
\begin{enumerate}[leftmargin=*, label={\arabic*.}]
    \item \textbf{Interest tag weight learning}: Dynamically adjust preferences based on users’ actual clicks and completion rates, weighting high-interest tags.  
    \item \textbf{Optimization of recommendation time windows}: Recommend lightweight content during fragmented time periods (e.g., leisure time in the morning or evening) based on user profiles.  
    \item \textbf{Scenario-based recommendation}: Increase recommendations for scenario-specific content (e.g., dietary health combined with home life).  
\end{enumerate}  

\textbf{3.5 Video Quality and Production Optimization}
\begin{enumerate}[leftmargin=*, label={\arabic*.}]
    \item \textbf{Enhance title appeal}:  
    \begin{itemize}[leftmargin=*, label={-}]
        \item Avoid clickbait and attract users with substantive content.  
    \end{itemize}

    \item \textbf{Optimize information density}:  
    \begin{itemize}[leftmargin=*, label={-}]
        \item Increase information content or highlight key points to make content compact and efficient.  
    \end{itemize}

    \item \textbf{Improve video quality}:  
    \begin{itemize}[leftmargin=*, label={-}]
        \item Ensure clear visuals and simple composition to adapt to the visual needs of elderly users.  
    \end{itemize}
\end{enumerate}  

\textbf{4. Conclusion}
Through the aforementioned optimization strategies, the recommendation system can better meet the needs of the health and wellness enthusiast group. The key lies in adjusting tag weights, optimizing content duration and opening appeal, and avoiding negative content to enhance user experience.  

The ultimate goals of the recommendation system should be:  
\begin{itemize}[leftmargin=*, label={-}]
    \item More accurate interest-based recommendations.  
    \item Higher content completion rates.  
    \item Stronger user interaction experiences.  
\end{itemize}

\newpage
\subsection{A.9 Detailed Visualization of UFM Agent}

In the main text, due to the excessive length of the reasoning output by GPT-4o, we placed the GPT-4o output here.

\noindent\fbox{\parbox{\linewidth}{

\textless think\textgreater 

Let me think. First, let's analyze the user's profile and interests. 

The user is a 54-year-old male, a public servant, and part of the small-town middle-aged demographic. His interests span a wide range, including social issues, entertainment, finance, technology, and more. Notably, he has shown interest in topics like "international society," "positive energy," "aviation and aerospace,"\textbf{ and "sports like kung fu and martial arts." These interests suggest a curiosity about extraordinary feats and global events.}

Now, let's consider the video content. The video is about an extreme challenge where someone jumps off the Burj Khalifa, an 800-meter-tall skyscraper. This aligns with the user's interest in extraordinary and daring activities, as well as international landmarks. The video length is 22 seconds, which is relatively short and should not be a barrier to completion.

Next, let's examine the user's viewing history. The user tends to watch short videos, often completing them if they align with his interests. For example, he watched a video about "a man drinking quietly on a train" (9 seconds, 50\% play rate) and "a helpful woman assisting a mother with children" (13 seconds, 50\% play rate). However, he skipped longer videos or those with less engaging content, such as "a singer being eliminated from a show" (51 seconds, 12\% play rate) and "a map trivia video" (70 seconds, 14\% play rate). This suggests that the user is selective and prefers concise, engaging, and relevant content.

From a value perspective, the video does not seem to contain negative or vulgar elements, and it likely showcases a positive or awe-inspiring achievement. This aligns with the user's tolerance for "positive energy" content.

Considering algorithmic factors, the recommendation seems accurate, as the video matches the user's interest in international landmarks and extraordinary activities. There is no indication of repetitive or irrelevant recommendations here.

Finally, regarding user-driven factors, there is no evidence of operational errors or lack of intent to watch videos at this time.

\textbf{In summary, the video aligns well with the user's interests}, is short enough to maintain attention, and does not contain any apparent flaws or mismatches. The user is likely to watch this video without skipping. 

\textless/think\textgreater

\textless answer\textgreater A. The user will not fast-skip this video. \textless/answer\textgreater

}}

\newpage

\subsection{Other Prompts}

\noindent\fbox{\parbox{\linewidth}{
\textbf{Prompt for generating SFT Data:}

This is a User profile: 

Age: \textbf{\{user age\}}

Gender: \textbf{\{user gender\}}

Occupation: \textbf{\{user occupation\}}

Other information \textbf{\{user other information\}}

User’s video interest tags: \textbf{\{interest tags\}}

history: \textbf{\{user history\}}

This is the information of the currently watched video:
Title: \textbf{\{watched video title\}}

Video duration: \textbf{\{video length\}}

Video vision info \textbf{\{video input\}}

The reason why the user dislikes this video is that \textbf{\{User feedback\}}

Determine which of the following reasons this reason belongs to. Additionally, based on user profile, viewing history, and watched video information, provide a chain-of-thought analysis explaining why the user dislikes this video.

\textbf{B:} Content-driven fast-skipping (due to objective content flaws) e.g., vulgar content, clickbait titles, physiologically discomforting visuals (e.g., bloody scenes, unpleasant creatures like snakes or centipedes).

\textbf{C: }Algorithm-driven fast-skipping (due to recommendation ineffectiveness) e.g., inaccurate user profiling, repetitive recommendations, insufficient diversity.

\textbf{D:} User-driven fast-skipping (arising from individual user actions)
such as operational errors and a lack of viewing intent at the current time.

You can provide the analysis from the aspects of values, interest matching, tolerance for negative energy, algorithm matching, content quality, and the user's own reasons. 

It's encouraged to include self-reflection or verification in the reasoning process.

\textbf{Provide your chain of thought reasoning process} between the  \textless think\textgreater  \textless /think\textgreater tags, and then give your final answer between the  \textless answer\textgreater  \textless /answer\textgreater tags.

}}

\end{document}